\documentclass{emulateapj}

\def\fnsiz{\footnotesize}
\newcommand{\as}{\mbox{$''$}}
\newcommand{\apsis}{{\sl Apsis\/}}
\newcommand{\aXe}{{\sl aXe\/}}
\newcommand{\bi}{\mbox{$B_{435} - i_{775}$}}
\newcommand{\calacs}{{\sl CALACS\/}}
\newcommand{\etal}{et~al.\/}
\newcommand{\EW}{\mbox{$EW$}}
\newcommand{\fion}[2]{\mbox{\rm [{#1}\thinspace{\footnotesize {#2}}]}} 
\newcommand{\Fline}{\mbox{$F_{\rm line}$}}
\newcommand{\Hline}[1]{\mbox{H{\footnotesize {#1}}}}
\newcommand{\Halpha}{\Hline{\mbox{$\alpha$}}}
\newcommand{\Hbeta}{\Hline{\mbox{$\beta$}}}
\newcommand{\Hgamma}{\Hline{\mbox{$\gamma$}}}
\newcommand{\iz}{\mbox{$i_{775}-z_{850}$}}
\newcommand{\HI}{{\sc Hi}}
\newcommand{\HII}{\mbox {\sc H ii}}
\newcommand{\kms}{\mbox{km\thinspace s$^{-1}$}}
\newcommand{\Lya}{\mbox{Ly{\fnsiz$\alpha$}}}

\newcommand{\SEx}{{\sl SExtractor\/}}
\newcommand{\zphot}{\mbox{$z_{\rm phot}$}}
\newcommand{\zspec}{\mbox{$z_{\rm spec}$}}
\newcommand{\zgrism}{\mbox{$z_{\rm grism}$}}

\slugcomment{AJ accepted}

\shorttitle{Emission-line sources in ACS grism data}
\shortauthors{Meurer et al.}

\begin{document}

\title{Automated selection and characterization of emission-line sources in ACS
WFC grism data}

\author{Gerhardt R.\ Meurer\altaffilmark{1}, Z.I.\
Tsvetanov\altaffilmark{1}, C.\ Gronwall\altaffilmark{2}, P.\
Capak\altaffilmark{3}, J.P.\ Blakeslee\altaffilmark{4}, N.\
Ben{\'{\i}}tez\altaffilmark{5}, H.C.\ Ford\altaffilmark{1}, G.D.\
Illingworth\altaffilmark{6}, L.D.\ Bradley\altaffilmark{1}, 
N.\ Pirzkal\altaffilmark{7}, J.\ Walsh\altaffilmark{8}, 
R.J.\ Bouwens\altaffilmark{6}, and S.\ Srinivasan\altaffilmark{1}}

\altaffiltext{1}{Department of Physics and Astronomy, Johns Hopkins
                 University, 3400 North Charles Street, Baltimore, MD
                 21218.}
\altaffiltext{2}{Department of Astronomy and Astrophysics, The
                 Pennsylvania State University, 525 Davey Lab, 
                 University Park, PA 16802.}
\altaffiltext{3}{Department of Astronomy, MS 105-24, California 
                 Institute of Technology, 1200 East California
                 Boulevard, Pasadena, CA 91125.}
\altaffiltext{4}{Department of Physics and Astronomy, Washington State 
                 University, Pullman, WA 99164.}
\altaffiltext{5}{Inst. Astrof\'{\i}sica de Andaluc\'{\i}a (CSIC), 
                 Camino Bajo de Hu\'{e}tor, 24, Granada 18008, Spain.}
\altaffiltext{6}{UCO/Lick Observatory, University of California, Santa
                 Cruz, CA 95064.}
\altaffiltext{7}{STScI, 3700 San Martin Drive, Baltimore, MD 21218.}
\altaffiltext{8}{ESO/ST-ECF, Karl-Schwarzschild-Strasse 2, D-85748 
                Garching bei M{\"u}nchen, Germany}

\begin{abstract}
We present complimentary techniques to find emission-line targets and
measure their properties in a semi-automated fashion from grism
observations obtained with the Advanced Camera for Surveys aboard the
{\em Hubble Space Telescope\/} ({\em HST\/}).  The first technique is to find all likely
sources in a direct image, extract their spectra and search them for
emission lines.  The second method is to look for emission-line sources
as compact structures in an unsharp masked version of the grism image.
Using these methods we identify 46 emission-line targets in the Hubble
Deep Field North using a modest (three orbit) expenditure of {\em HST\/}
observing time.  Grism spectroscopy is a powerful tool for
efficiently identifying interesting low luminosity, moderate redshift
emission-line field galaxies.  The sources found here have a median $i$
band (F775W) flux 1.5 mag
fainter than the spectroscopic redshift catalog of Cohen et al. (2000).
They have redshift $z \leq 1.42$, high equivalent
widths (typically $EW > 100$\AA), and are usually less luminous than the
characteristic luminosity ($L^\star$) at the same redshift.
The chief
obstacle in interpreting the results is line identification, since the
majority of sources have a single emission line and the spectral
resolution is low.  Photometric redshifts are useful for providing a
first guess redshift.  However, even at the depth of the
state-of-the-art ground-based and {\em HST\/} data used here, photometric errors can result
in uncertainties in line identifications, especially for sources with
$i$ magnitudes fainter than 24.5 ABmag.  Reliable line
identification for the faintest emission-line galaxies requires
additional ground-based spectroscopy for confirmation.  Of particular
concern are the faint high $EW$ \fion{O}{II} emitters which could
represent a strongly evolving galaxy population if the possibility that
they are mis-identified lower redshift interlopers can be ruled out.
\end{abstract}


\section{Introduction }\label{s:intro}


One of the perennial problems in modern astrophysics is measuring
spectral information, especially redshifts, of the most distant and
hence, faintest sources in the universe.  With the {\it Hubble Space
Telescope\/} ({\em HST\/}) and the Advanced Camera for Surveys' (ACS) Wide Field
Camera (WFC) astronomers can now obtain reliable broad band photometry
down to ABmag $\sim 27$ across the optical portion of the spectrum with
a modest expenditure of telescope time \citep[e.g.][]{b04}.  The
instrument is able to image even deeper as demonstrated by the Hubble
Ultra Deep Field \citet[HUDF][]{hudf06} which has a $S/N = 10$ limit of
$\sim$29.9 ABmag for point sources with the F775W filter \citep{bibf06};
in the same field and using the same filter the $S/N = 10$ detection
limit for extended sources within an aperture having diameter of 0.4\as\
and 0.5\as\ is 29.4 and 29.0 ABmag, respectively \citep{hudf06,coe06a}.
However, our limit for obtaining reliable spectroscopy is much
brighter. Spectra to ABmag $\approx$ 24 are difficult to obtain even
with the largest ground-based telescopes \citep[e.g.\
][]{gdds1,cbhcs04,c00}.  At issue is the domination of the sky over the
signal from astronomical sources observed from the ground at these faint
magnitudes.  The background seen by {\em HST\/} is orders of magnitude weaker.
However, slit survey spectroscopy with the now defunct STIS and FOS
spectrographs was not feasible due to the telescope's small aperture and
minuscule projected slit widths.  Alternatively, multi-object slitless
spectroscopy is and has been available with {\it HST\/} using a variety
of instruments (FOC, STIS, NICMOS).  All three cameras of ACS also have
dispersing elements.  The combination of the ACS G800L grism and the WFC is
particularly noteworthy since it provides the widest field and highest
throughput of all the slitless options on {\it HST\/}, allowing deep
low-resolution spectroscopy with modest expenditure of telescope time.
For example the three orbit integration of the Hubble Deep Field North
(HDFN) that we discuss here provide $S/N \sim 10$ spectra in the
continuum of sources having ABmag = 25.1 at a mean wavelength $\lambda
\approx 8000$\AA.

While this major new capability is welcomed, grism data are difficult to
work with.  The broad spectral coverage of G800L \citep[$\lambda = 5600
- 9900$\AA\ at 25\%\ of peak spectral throughput,][]{wp04} results in a
background count rate that is high compared to other WFC filters, and
indicates that wavelength variations in the flat-field are a major
concern.  While the grism transmits most of its light in the first
order, light from other orders of bright sources can also be seen.  The
orientation of the G800L filter and the strong geometric distortion of
the ACS \citep{acs-distort02} results in spectral traces that are skewed
relative to the CCD pixel grid at an angle that varies across the field
and a wavelength calibration that also is field dependent.  The
spatially varying distortion makes combining dithered G800L images
difficult, especially if there are large offsets or roll-angle
variations.  Fortunately astronomers at the Space Telescope European
Coordinating Facility (ST-ECF) have provided calibrations of the G800L
grism \citep{ppw03,wp04} as well as the \aXe\ software package for
extracting and processing slitless spectra \citep{ppd01} which relieves
many of the pains of dealing with grism data.

The aim of this paper is to assess techniques to process grism data and
identify Emission Line Galaxies (ELGs).  ELGs are particularly
interesting since they mark the location of AGN or intense star
formation and hence, cosmic evolution.  For example, some of the most
distant galaxies in the universe are \Lya\ emission sources at $z \sim
6.6$ \citep{kodaira03,taniguchi05,kashikawa06}.  In addition, sharp
narrow emission features should be one of the easiest spectral
signatures to find in grism data.  Since this is primarily a techniques
paper, our science analysis is relatively light, and includes basic
comparisons of the key measurable properties of the ELGs (luminosity,
equivalent width, color) with other galaxy samples at a similar
redshift.  This analysis is sufficient to show that the grism provides
an efficient means for selecting statistically significant samples out
to $z \sim\ 1.5$.  This study is similar to \citet{grapes1} which
provides a detailed description of how the GRism ACS Program for
Extragalactic Science (GRAPES) collaboration have reduced their G800L
observations of the HUDF and extracted source spectra.
\citet{egrapes06a} present their technique for identifying ELGs and the
resultant catalog of sources found, while \citet{egrapes06b} discuss the
morphology of these ELGs.  Here we discuss WFC grism processing methods
and tools that were developed by the ACS Science Team largely
independently from the ST-ECF and GRAPES efforts and optimized for the
discovery of ELGs.  We apply these methods to observations of the HDFN.
A brief summary of this work was presented at the 2005 {\it HST\/}
Calibration Workshop \citep{meurer06}.

In Sec.~\ref{s:over} we compare our data processing to that of the GRAPES
collaboration and describe in detail our data and its processing.  In
Sec.~\ref{s:felg} we present two methods for finding ELGs, as well as our
methods for assigning line identifications.  Sec.~\ref{s:results} presents
our results including a list of all ELGs found, and an initial
assessment of the statistical properties of the galaxies found.  In
Sec.~\ref{s:compare} we compare our redshifts with other observations of
the HDFN and determine the redshift accuracy of the grism.  Finally, in
Sec.~\ref{s:disc} we summarize our results, discuss the benefits of using
the grism and the additional requirements for obtaining useful redshifts
of ELGs.

\section{Methods and data }\label{s:over}

Here we provide an overview of our image processing and object
extraction and note how it differs from that of the GRAPES collaboration
\citep{grapes1}.  Following that, we describe the processing of the HDFN
data in detail.

\subsection{Comparison of image processing techniques}

The GRAPES team do minimum processing of their images before extracting
spectra.  Like us, they rely on the STScI calibration pipeline \calacs\
\citep{hack1999}, as implemented by the STScI archive to do most of the
basic CCD processing consisting of overscan subtraction, bias
subtraction, dark subtraction, and gain correction. These steps are
performed using the best available reference files as implemented by the
STScI Archive.  The flat-field employed by \calacs\ for G800L images is
a unity flat, so in effect no pixel-to-pixel flat-fielding is done.  The
GRAPES team subtract a scaled super-sky frame from each image to remove
the sky background (where the scaling is to object free regions of the
image).  The G800L image shifts are determined from {\sl
  MULTIDRIZZLE}-processed short exposures obtained through a broad-band
filter at the start of each orbit.  The {\sl MULTIDRIZZLE\/} task is
also used to produce geometrically-corrected G800L images, but they are
used only to identify the cosmic rays, not for spectral extractions.
Instead, the extractions are performed by \aXe\ on the individual frames
after sky subtraction and masking of the cosmic rays.  The spectra are
co-added and the flat-fielding is performed at this stage by calculating
the effective wavelength of the light falling on each pixel and
interpolating between a series of broad- and narrow-band flats
\citep{wp04} to determine the flat field correction most appropriate for
that pixel.

Our approach differs in a few key ways.  First, we flat-field our G800L
images with the flight flat-field for the F814W filter.  We then process
the data with the GTO science pipeline \apsis\ \citep{bambm03} to make
cosmic-ray rejected, aligned, combined, and geometrically corrected images.
The result is one final G800L image which we use to extract spectra
and all spectroscopic quantities.  There
are several advantages of this approach. Application of the F814W flat
cosmetically improves the images by largely removing most small-scale
CCD blemishes.  In addition it produces flatter images, reducing the rms
amplitude of the sky background over large scales ($\gtrsim 75$ pixels)
by a factor of two, as determined from application of the flat-field to
the super-sky frames used by \citet{grapes1}. Since we forgo the use of
aXe's $\lambda$ dependent flat-field fit, the flux scale varies
throughout the field of our images by up to $\sim$10\% \citep{wp04}.
Flat-field images from WFC show numerous dark blemishes which are more
apparent with decreasing $\lambda$ \citep{bhm01}.  Because of the broad
spectral response of the grism, blemishes are inaccurately removed with
the F814W flat. For example, if the blue end of the spectrum of a compact source falls
on a blemish, it will not be completely removed by flat
fielding and result in a spurious absorption feature.  However, this is
not a major concern here since we are concerned with emission lines
rather than absorption features.  This is also less of a problem for
extended sources ($\gtrsim 10$ pixels): since many wavelengths
contribute to each pixel in the spectrum one can no longer assume that a
single wavelength dominates, and hence our flat-fielding technique
should be sufficient in these instances.  Using small dithers can also
mitigate against this happening.  Combining geometrically corrected
dithered data can work as long as there are no roll angle variations and
the dithers are all within $\sim 6$\as.  Then, the $\lambda$ scales of
the first order spectra will align to within 0.025\as\ ($\sim$ 0.5 WFC
pixels) across the WFC field.  Geometric correction has the advantage
that it removes much of (but not all of) the spatial variation in the
$\lambda$ calibration with the dispersion remaining nearly constant
within each spectrum. A major advantage of our approach is that the
geometrically corrected spectra are nearly horizontal.  Over a spectral
length of 75 pixels, we calculate a slope of 0.03 pixels averaged over
the geometrically corrected image (maximum slope of 0.99 pixels), while
the average slope in the raw images is 2.54 pixels (maximum 3.43
pixels).  Horizontal spectra are easier to extract and analyze using a
variety of tools.  The orientation also allows simple filtering to
remove cross-dispersion structure and isolate emission lines (e.g.\
Sec.~\ref{ss:grsel}).  Finally, \apsis\ processing of the images provides
excellent cosmic-ray and hot-pixel removal and removes a small amplifier
step (typically having an amplitude of a few electrons) often seen in
WFC images.

\subsection{The data}\label{s:data}

The HDFN field (RA = $12^{\rm h}36^{\rm m}47.11^{\rm s}$, Dec =
$+62^\circ 13' 11.9''$) was observed by the ACS science team (program number
9301) for 2 orbits in the F775W ($i_{775}$) filter, and 3 orbits each
with the F850LP ($z_{850}$) filter and G800L grism as summarized in
Table~\ref{t:obs}.  Two exposures per orbit were obtained in order to
facilitate cosmic-ray removal, and the telescope was dithered by 1 pixel
in each axis between orbits.  The individual \calacs\ processed G800L
``FLT'' frames were divided by the standard F814W flat-field image.
Fine alignment of the individual images was performed with \apsis.
\apsis\ combined the individual exposures to make a single aligned image
for each filter, F775W, F850LP, and G800L, using a spatial sampling of
0.05\as\ per pixel, as well as a detection image
that is the inverse sky variance weighted sum of the F775W and F850LP
images.  Because most objects in grism images are rather elongated and
faint, \apsis\ could not accurately register the G800L FLT images and
determine the offsets.  Instead it employed default shifts determined
from the positions which are stored in the image headers.  We used a
modified version of \apsis\ to check the shifts in the G800L ``CRJ''
images from the STScI \calacs\ pipeline - these are pairs of images
combined to form a single cosmic-ray rejected image.  There are three
CRJ images for this dataset (one for each G800L orbit).
\apsis\ matched 8-10 zero-order images of bright sources compared to the
reference G800L CRJ image, yielding average shifts accurate to
$\sim$0.05 pixels in each axis.  The resultant shifts matched the
default shifts to 0.1 pixels.  The final \apsis\ drizzle cycle was done
to an output scale of 0.05\as\ per pixel with interpolation performed
using a Lanczos3 kernel.  The Lanczos3 function, defined by
\citet{mei05}, is a damped-sinc function. Application of it during
drizzling results in better preservation of the noise characteristics
and spatial resolution of the data than the standard linear (square)
interpolation kernel \citep{mei05}.  The FWHM resolution of the final
F775W and F850LP images was measured from direct measurement of stellar
radial profiles and reported in Table~\ref{t:obs}.  For the G800L image
we measure the spectral resolution from cross-dispersion cuts (five column sums) of
the first order spectra of stars.  These were fitted with a Gaussian
profile, and the resolution taken as the average FWHM of the fits.  The
resulting resolution of 2.1 pixels corresponds to $R \equiv \lambda /
\Delta \lambda \approx 90$ at $\lambda = 8500$\AA\ at the center of the
field.

We used the flux calibration curve given in \citet{wp04} to convert
spectra to flux units.  As noted above, there is a $\sim 10$\%\
variation in the flux scale across the field when applied to data
processed outside of the standard \aXe\ extractions from FLT frames.  We
employed a $\lambda$ calibration determined from WFC G800L images of
Wolf-Rayet stars (which have strong bright emission lines) that were
observed so as to fall on various positions on the WFC detector.  The
data and measurement techniques employed are identical to those used by
\citet{ppw03}, but applied to the calibration data after drizzling them
onto a rectified pixel grid with an output pixel scale of 0.05\arcsec,
the same as our data. The resultant $\lambda$ calibration is given as a
quadratic polynomial as a function of column offset from the
geometrically corrected direct image, with the polynomial's coefficients
varying quadratically with position in the corrected direct images.

\section{Finding emission-line galaxies }\label{s:felg}

We have developed two methods for the semi-automated identification and
classification of emission-line galaxies which we detail here.  Here we
define the term ``emission-line galaxy'' (ELG) to be a galaxy having
line emission detected in our grism images. In principle, line emission
in an ELG may be dispersed evenly throughout the galaxy.  In practice it
is usually confined to a small region, such as a nucleus or a knot.  We
use the term ``emission-line source'' ELS to denote a source of line
emission that is distinct in position and wavelength.  Effectively an
ELS can be isolated as a distinct source in the grism image.  Hence, an
ELG with two knots each with only one detected line has two ELSs.
However, if each of its knots has two distinct lines then there are four
ELSs in the system.

\subsection{A: aXe selection }\label{ss:axesel}

Method A (for \aXe) is very similar to that employed by the GRAPES team
\citep{grapes1,egrapes06a}.  The extractions are done using \aXe\ with
the calibrations discussed above (Sec.~\ref{s:data}) encoded into its
configuration file.  The extractions are done from the \apsis\ processed
G800L image which has a low order sky background subtracted from it.  No
additional sky subtraction was performed.  The \aXe\ input catalog was
derived from a \SEx\ \citep{ba96} catalog of the detection image.  Since
we are starting with spectra of entire galaxies (and stars), any line
emission we find can be attributed to the galaxy as a whole (making it
an ELG) but not localized further.  However, multiple ELSs can be
discerned within an ELG if there is more than one emission line.  We
configured \aXe\ to extract spectra down to the detection limit $\sim
28.7$ ABmag in the detection image.  We go this faint to maximize the
chance that we find faint ``pure'' ELGs - galaxies that have one
emission line and no continuum.  The grism observations have a similar
exposure time and system throughput as the direct images, hence a pure
ELG will have similar count rates in the direct and grism images.
Taking the differing exposure times and the rms noise level of the sky
into account we calculate that the $S/N$ ratio of a pure ELG would be
similar in our detection and grism images.  We caution that this
condition may not hold for other datasets.

The detection image catalog was processed to remove sources with $m >
28.7$ ABmag (too faint) and semi-minor axis size $b < 0.8$ pixels (too
small, most likely image defects).  \aXe\ was configured to set the extraction
aperture equal to 2.5 times the projected semi-major axis size, $a$ of
the objects.  However, first we reset the size of sources having $a < 2$
pixels or $25.4 < m \leq 28.7$ ABmag to $a=2$, $b=2$ pixels.  Hence our
smallest extraction aperture is 5 pixels wide.  The final step before
extracting the spectra was to mask the area within 8 pixels of the edges
of the CCD chips in the grism image due to the number of false positives
we found in preliminary runs with our code.

One-dimensional flux-calibrated spectra were extracted with \aXe.
Automatic identification of ``interesting'' targets (ELG candidates) was
performed by subtracting a smooth baseline spectrum and finding the
sources with residuals having a peak $S/N \geq 4$.  The baseline was
constructed by median filtering the spectrum and then boxcar smoothing
it, using a filter size of 19 pixels in both steps.  The spectrum of
each candidate was displayed and classified as either (a) an ELG, in
which case one or more Gaussian components were fitted to the peak(s) in the
spectrum; (b) a ``star'', that is a source with strong broad absorption
lines - our algorithm often mistakenly identifies the peaks
between the absorption features as emission lines \citep[the absorption lines
sources are typically late type M or K stars although we also found
the two supernovae discussed by][]{acs_sne03}, or (c) a spurious
source.  The sources classified in this manner are discussed in
Sec.~\ref{s:results}.

\subsection{B: Blind grism image selection }\label{ss:grsel}

Method B (for blind selection) starts with the grism image and is
designed to identify all detectable ELSs within galaxies.  It is a
``blind'' selection in the sense that we do not require the {\em a
priori\/} knowledge of source positions to find the ELSs.  We find this
to be very useful for two reasons.  First, the ELSs we find are often
confined to nuclei, off-center starbursts or strong \HII\ regions.
Normal \aXe\ extraction, as in method A, may dilute the line signal with
``unnecessary'' continuum flux or report the incorrect $\lambda$ for the
line if it results from an off-center knot.  This is because in \aXe\
extractions, the $\lambda$ of each pixel depends on its offset from the
major axis in the direct image; $\lambda$ errors may then occur for
knots offset from the major axis.  Since the flux scale depends on
$\lambda$, a flux error will also result.  Second, method A can not find
all possible pure emission-line sources.  A pure ELG that emits at
5600\AA\ $\lesssim \lambda \lesssim$ 7100\AA\ will be invisible in our
direct images since the the filters we employ do not have significant
sensitivity at these wavelengths.  However, the grism does, and hence we
may still hope to find such sources, if they exist, in our grism images.

Processing starts with high-pass filtering both the grism image and the
detection image.  This is accomplished by smoothing with a $13 \times 3$
median filter and then subtracting this smoothed image from the
original. The long axis of the filter is parallel to the image rows,
that is, very nearly parallel to the dispersion direction.  The
filtering effectively removes most of the continuum in the grism image
and much of the low frequency spatial structure of the detection image,
leaving compact ELSs in the grism image and galaxy nuclei, bars and
knots in the direct images.  The high-pass filtered grism images also
contain the zero-order images, offset by $\sim -115$ columns from the
direct images.  These could be mistaken for ELSs.  So, before searching
for emission-line candidates we mask those that could contaminate our
results. This is done by determining a linear coordinate
transformation\footnote{$X_{\rm out} = a + bX_{\rm in} + cY_{\rm in};
Y_{\rm out} = d + eX_{\rm in} + fY_{\rm in}$} between the detection
image positions and that of the the zeroth-order images as well as a
mean flux ratio.  In terms of count rate, objects in the F775W (F850LP)
image are on average 32 (21) times brighter than their zeroth-order
counterparts in the grism image.  We apply the appropriate flux ratio to
the detection image to locate pixels that would have a flux equal to or
greater than the sky noise in their zeroth order. The coordinate
transformation is used to determine their location in the grism image.
This pixel distribution is grown by a radius of three pixels by
convolving it with a circular top-hat function.  The resultant masked
pixels are set to 0.0 in the filtered grism image.  The total usable
area of the image is then 1659972
pixels or 11.83 arcmin$^2$. Using this masking,
about 60 spurious sources are excluded from the source catalogs
(discussed below), while only 0.13\%\ of the otherwise good area of the
image is masked out.   Hence, the masking is very efficient at removing
spurious sources yet unlikely to remove many real ELSs
from the grism image.  Zero-order images may survive near the image
edges where the direct image falls outside the field of view of our
detection image.  This condition is easily tested.  Stars and very
compact sources also remain in the high-pass filtered images because
they are sharper than the smoothing box cross-dispersion width.
However, they are easily recognized and flagged in the classification
stage.

We use \SEx\ to find sources in the masked and filtered grism image.  By
experimentation, we found that setting \SEx\ parameters DETECT\_THRESH
and ANALYSIS\_THRESH to 1.15 and DETECT\_MINAREA to 3 was sufficient to
find most obvious compact line emitters visible by eye without
introducing large numbers of spurious detections.  We removed from this
catalog sources with output parameters ELONGATION (axial ratio) greater
than 2.5 (likely spectral continuum residuals), B\_IMAGE less than 0.4
or FWHM\_IMAGE less than 1 pixel (likely residual cosmic rays or hot
pixels), or FWHM\_IMAGE greater than 7 pixels (spurious since sources
this large should have be missing from the high-pass filtered images).
For each source we extract a region extending from $-150$ to $+10$
columns from its position in the grism image and having an extraction
width $\Delta y$ rows equal to 1.25 times its size projected onto the
cross-dispersion axis:
\begin{equation}
\Delta y = 1.25 \sqrt{(a \sin \theta)^2 + (b \cos \theta)^2}.
\end{equation}
Here $a,b$ are the semi - major, minor axes A\_IMAGE, B\_IMAGE from
\SEx\, and $\theta$ is the position angle measured counter-clockwise
from the $+x$ axis defined as having constant row number in the pixel
grid and directed towards increasing column number.  The $+x$ axis is
close to, but not exactly the dispersion axis directed towards
increasing wavelength.  We set a minimum $\Delta y = 5$ rows for the
extraction.  This region is extracted from both the grism and direct
filtered images, then the rows are summed to make 1D-cuts.  The region
outside of a 13 pixel box centered on the emission line is set to 0.0 in
the grism cut to isolate the ELS.  The grism and direct cuts are then
cross-correlated to determine the $x$ offset between the ELS and sources
in the direct image.  If the cross-correlation peak corresponds to the
correct source in the direct image, then the $x$ offset yields the
source position in the direct images and hence a preliminary estimate of
the wavelength of the line.

Final measurements of the emission-line quantities are obtained from 1D
spectra of each knot extracted with \aXe\ using the cross-correlation
determined position, and an extraction aperture of 5 pixels.  The
emission-line properties are measured with Gaussian fits as in
Sec.~\ref{ss:axesel}.  However, we use a peak $S/N = 3$ cut, lower than
that employed by method A, since we find that method B can (usually)
reliably find ELSs at this low of a significance level.  Comparison of
the fluxes of 19 single knot ELSs found by both methods show that lines
are on average 0.04 dex brighter (with a dispersion of 0.14 dex) when
measured with method A compared to B.  We consider this not to be a
significant difference.

\begin{figure}
\epsscale{0.5}
\plotone{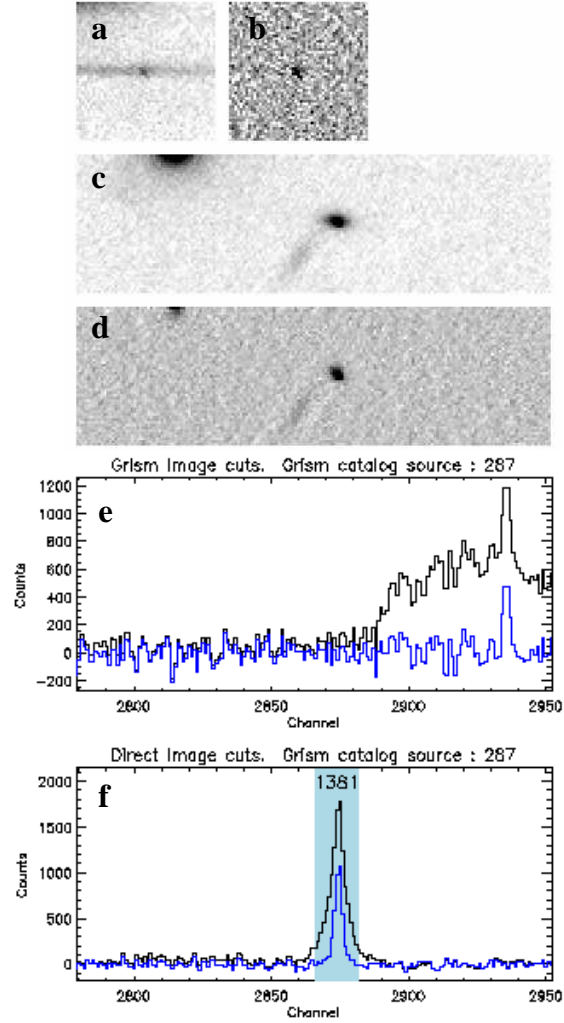}
\caption[]{\small{Steps in the processing of the grism and detection image for
  finding ELGs in the grism images and measuring their
  properties. Panels a and b show a 50$\times$50 pixel cutout of the grism
  image before and after (respectively) subtracting a 13$\times$3 pixel median
  filtered version of the image (high-pass filtering).  These cutouts
  are centered on an emission-line candidate selected from the high-pass
  filtered grism image.  Panels c and d show cutouts of the direct image
  before and after high-pass filtering.  The width of the cutout is
  selected to include the full range over which the direct image
  counterpart to the source seen in panel b may reside. Panel e shows
  the 1D spectra made by extracting and summing five rows centered on
  the emission line from the grism image before (black line) and after
  (blue line) high-pass filtering.  Panel f shows the same thing for
  the 1D cuts through the direct image.  The shaded regions show the
  pixels in the collapsed region that belong to an object found by \SEx\
  (the working object identication number is shown).
  \label{f:grexamp}}}
\end{figure}

Figure~\ref{f:grexamp} shows an example of the image manipulation and
detection process for this method.  Since there are a number of ways
this method can produce spurious results, this technique is applied in
an interactive environment.  For each candidate ELS,
the cutouts of the grism and direct images are examined, and used to
assess whether it corresponds to a blemish in one of the images or a
star (like method A, this technique is also adept at finding compact
broad absorption-line sources).  Line plots of the 1D cuts and the
cross-correlation spectrum are produced, as is the auto-correlation of
the high-pass filtered grism spectrum with itself.  The peaks in the
cross-correlation are fitted with Gaussian profiles until the residuals
have no peaks with $S/N \lesssim 3$.  If there is more than one
Gaussian component in the fit, the correct match is interactively
selected using the 2D cutouts and line plots as a guide - an ELS
typically corresponds to a high-surface brightness compact nucleus or
knot with its 2D line image resembling the high-pass filtered grism
image in size and orientation.  Spurious peaks in the cross-correlation
typically can be identified (by eye) as having the wrong cross-dispersion position or
not corresponding to a nucleus or knot.  There remain cases, however, of
more than one plausible direct counterpart to the ELS.  This could be
due to multiple knots in the direct image, or completely separate
sources.  We flag these ambiguous cases.  The sources classified by this
technique are discussed in Sec.~\ref{s:results}.

\subsection{Line identification and redshift}

Two emission lines are found in seven ELGs, allowing line identification
and redshifts to be determined using the ratio of observed wavelengths,
which is invariant with redshift.  In three cases the two lines are
rather close and clearly correspond to \Hbeta\ and the
\fion{O}{III}$\lambda$4959,5007\AA\ doublet, which is blended at the
grism's resolution.  In two cases the ratio of wavelengths indicates
that the lines are \Halpha\ and \fion{O}{III}.  Note that one must be
careful with this technique since $\lambda_{\rm H\alpha}/\lambda_{\rm
[OIII]} = 1.3138$ is close to $\lambda_{\rm H\beta}/\lambda_{\rm [OII]}
= 1.3041$.  A one pixel uncertainty {\rm in both\/} line wavelengths
could result an ambiguous line identification over the redshift range of
interest.  Our adopted identification in these two cases corresponds to
previous spectroscopic and photometric $z$ estimates (see below),
indicating that our identifications are correct.  In one case we
identify the lines as \Hbeta\ and \Hgamma\ at $z = 0.947$, while in
the last case we identify \fion{O}{II} and \fion{Ne}{III}3869\AA.

The vast majority (39/46) of ELGs found contain only a single detected
line.  Identification of this line is a primary, but difficult, task.
The line may well be a blend at the low resolution of our data (e.g.\
\Halpha\ and \fion{N}{II}, the \fion{O}{III} doublet, and the
\fion{O}{II}$\lambda$3726,3728\AA\ doublet).  Similarly, the resolution
is not high enough to identify lines by profile shape (e.g.\ \Lya).  For
these sources, our approach is to use available redshifts as a first
guess to each source's redshift and then determine which line
identification agrees best with the guess.  

We use four sources for the redshift guesses, one spectroscopic source
and three photometric redshift sources.  Multiple estimates are used to
insure that all grism sources have at least one first guess redshift,
and as a consistency check to determine how resilient the line
identification is to the first guess source. In addition, the different
redshift sources represent different choices of strategies, and
investments in telescope time, labor and resources that may be employed
to obtain redshifts.  We have four redshift sources which we now list
with the number of matches to the 46 ELGs [in brackets]: (1)
spectroscopic redshifts (\zspec) from the compilation of the Hawaii
group \citep{cbhcs04} [22 matches];  (2) photometric redshifts (\zphot)
from \citet[][ hereafter C04]{cap04} who use terrestrial $U$, $B$, $V$,
$R$, $I$, $z$, and $HK$ photometry and derive \zphot\ estimates from
version 1.99 of the BPZ code described by \citet{bpz00} [37
matches]; (3) \citet[][ hereafter FLY99]{fly99} who use $U_{300}$,
$B_{450}$, $V_{606}$, and $I_{814}$ photometry which they measure from
the HDF data of \citet{hdf96} as well as ground-based $J$, $H$, and $K$
photometry from \citet{d98} to derive their \zphot\ estimates [25
matches]; and (4) our own \zphot\ estimates which we derive from the on-line
GOODS $B_{435}, V_{606}, i_{775}, z_{850}$ ACS photometry \citep[release
r1.1z,][]{goods04a} [43 matches].  For the latter we employed version
2.0-alpha of the BPZ code described by \citet{bpz00} which reports
results on up to three peaks in the $z$ probability distribution.  We
assume that the emission lines seen are one of the following: \Halpha,
the \fion{O}{III}4959,5007\AA\ doublet, the \fion{O}{II}3726,3729\AA\
doublet, or \Lya, for which we adopt rest-wavelengths in {\em vacuum\/}
of $\lambda_0 = 6564.6, 4996.5, 3728.7, 1216$\AA, respectively.  We
adopted intrinsic flux ratios of $F_{5007}/F_{4959} = 3.03$ for the
\fion{O}{III} doublet (set by quantum mechanics), and $F_{3728}/F_{3726}
= 1.3$, corresponding to an electron density $n_e = 100\, {\rm cm}^{-3}$
\citep[near the low density limit;][]{o89} for the \fion{O}{II} doublet.

\section{Results}\label{s:results}

\subsection{Data presentation}

Properties of the 53 ELSs in the 46 unique ELGs found by our techniques
are listed in Table~\ref{t:source}.  The first column gives the ELG
identification number. The sources are ordered by \zgrism, larger ELG
numbers indicate higher redshifts.  The second column gives a 
coordinate based name, mostly from the GOODSN r1.1z catalog.  Remaining
columns are explained in the table notes.  Postage stamp cutouts of the
ELGS taken from the detection image are given in Fig.~\ref{f:stamps}.
Two stamps are given for each source: one from the plain detection
image and the other from the high-pass filtered detection image.  The
former is ideal for identifying the field while the latter is well
suited to show knots and other sharp structures not always seen in the
plain detection image.  Comments on individual ELGs are given in
the appendix.

We found 32 ELSs in 30 ELGs using method A - two of these ELGs have two
emission lines detected with method A.  Method B reaped more ELSs - 49
in 39 unique ELGs.  Five of these have two lines detected with method B.
Two ELGs have two identified emission knots, and one has four
emission-line knots.  The object centers for the extracted spectra and
the method of detection is indicated in the stamps: squares mark sources
identified with method A and circles mark sources from method B. The line
identification corresponds to \fion{O}{III} in 26 cases, \fion{O}{II} in
13 cases, \Halpha\ in eight cases, \Hbeta\ in four cases, \Hgamma\ in
one case, and \fion{Ne}{III}$\lambda$3869\AA\ in one case.  There
were four candidate ``pure'' ELGs - apparent emission-line sources with
no direct counterpart.  However, careful examination of the grism image
and noise maps produced in the drizzling process show that they are all
probably imperfectly removed cosmic rays and hot pixels.

\subsection{Properties of grism-selected ELGs\label{ss:stats}}

Here we examine some basic measured properties of the sample.
These include line wavelength, flux, and equivalent width, as well as
continuum magnitudes, colors, and luminosities.  Our intent is to give
an overview of the properties of our sample, examine the extent to which
they are set by selection effects, and contrast them with other distant
galaxy samples.  It is not an analysis of the astrophysics of the ELGs,
nor do we address issues of cosmology and the evolution of galaxy
populations, since the emphasis of this paper is on how to find ELGs.

Figure~\ref{f:lamdist} compares the wavelength distribution of our
sample with the grism sensitivity curve.  The distributions from method
A and method B selection are shown separately.  Both distributions and
the curve peak at similar $\lambda$ and have a relatively long red tail.
However, there is deficit of detections of lines at $\lambda \lesssim
7000$\AA\ compared to the sensitivity curve.  One-sided
Kolmogorov-Smirnov tests yield probabilities of 7\%\ and 3\%\ that the
observed $\lambda$ distribution follows the grism sensitivity curves for
method A and B selection respectively.  So while the sensitivity curve
may play a significant role in determining which ELGs are selected,
other factors, including the volumes accessed by the individual lines,
the luminosity function and large scale structure also effect the
$\lambda$ distribution.

We compare our line flux distributions in Figure~\ref{f:flamdist}.  The observed
distributions are similar for the two selection techniques, although
there is a hint that method B is finding more faint lines than A.  The
peak in the distributions is at $\Fline \approx 4\times10^{-17}\, {\rm
erg\, cm^{-2}\, s^{-1}}$.  For comparison we show the \Fline\
distribution from the STIS Parallel Survey
\citep[SPS][]{teplitz03a,teplitz03b}.  The SPS is similar to our survey
in that it was obtained with {\it HST\/} using observations with a range of
exposure times (typically a few thousand seconds, i.e.\ similar to that
used here), albeit with an instrument, the Space Telescope, Imaging
Spectrograph, which has a much smaller collecting area and lower
throughput than our observing configuration.  We see that our
observations typically reach about four times deeper in line flux,
probably because of the improved sensitivity.  The peak in our \Fline\
distributions corresponds to a 10$\sigma$ detection within a 5 by 5
pixel pixel box for a line having $\lambda$ at the peak of the
sensitivity curve.  However, the $\lambda$ distribution is broad, and
many lines are detected off the peak in the sensitivity curve (cf.\
Fig.~\ref{f:lamdist}).  The distribution of measured signal-to-noise
ratio ($S/N$) is shown as insets in Fig.~\ref{f:flamdist}.  The $S/N$
histograms show that we start to lose lines at $S / N \lesssim 6$.  As
noted in sections \ref{ss:axesel} and \ref{ss:grsel} our initial source
selection requires a {\em peak\/} $S/N = 4$ and 3, respectively in 1D
spectra.  We are not finding weaker detections because here we are
showing $S/N$ measurements of {\em integrated\/} line flux within a 5 by
5 pixel box.  One could dig further down in flux by convolving the 1D
spectra with a matched filter, and then lowering the peak $S/N$
detection limit, although this would require weeding out more spurious
sources.

Figure~\ref{f:mhist} shows the $i_{775}$ magnitude distribution of the
ELGs in comparison to two other samples in this field - the Hawaii group
\zspec\ compilation of \citet{cbhcs04} and the \zphot\ catalog of C04.
Only the sources that match sources in our detection image are included
in the histograms to insure that the comparison covers the same area on
the sky.  For the grism sample, the median $i_{775}$ is 23.9 ABmag and
the 75th percentile is $i_{775} = 24.9$ ABmag.  These are significantly
fainter than $i_{775} = 23.1, 23.8$ ABmag respectively at these
percentiles for the Hawaii \zspec\ compilation.  Through expending
considerable effort and time (particularly with the Keck telescopes)
they were able to determine spectroscopic redshifts for over four times
as many sources (210) than the number of ELGs we find.  Nevertheless,
with a modest three orbit exposure we determined grism redshift
estimates for 23 sources with no previous \zspec\ estimates.

The faint apparent magnitudes correspond to low luminosities compared to
typical galaxies.  This is shown on in Fig.~\ref{f:mabshist} which plots
histograms of absolute magnitude in the rest-frame $B$ band, $M_B$ for
our ELG sample, divided by line identification.  We derived $M_B$ by
interpolating the broad band SED of the ELG host from the {\it HST\/}
GOODS optical photometry \citep{goods04a} to determine the apparent
ABmag at rest $\lambda_0 = 4297$\AA, the central $\lambda$ of the $B$
filter.  We used the apparent magnitude in the nearest filter with data
for cases where this $\lambda$ in the observed domain falls outside the
range of central wavelengths of the ACS filters used by GOODS.  The
absolute magnitude was then calculated for our adopted \zgrism\ assuming
$H_0 = 70$ km s$^{-1}$ Mpc$^{-1}$, $\Omega_M = 0.3$, and $\Omega_\Lambda
= 0.7$ \citep[following][]{cpt92}.  Table~\ref{t:elgstats} presents the
first quartile, median, and third quartile values of $M_B$ and \zgrism\
for our ELG sample, divided by emission line identification (ID).  We
use a thick gray line to mark the absolute magnitude corresponding to $m
= 27$ ABmag for an object at the redshift corresponding to the relevant
line being found at $\lambda=7500$\AA\ (the peak throughput of the
grism).  This gives a crude indication of the limiting magnitude for
each type of line emitter.  It is not an absolute limit, since less
luminous sources are still possible corresponding to finding somewhat
fainter apparent magnitudes (see Fig~\ref{f:mhist}) or bluer lines
(hence closer sources).  For comparison, we also mark the characteristic
absolute magnitude at the knee of the luminosity function, $M_B^\star$,
derived from various galaxy surveys in the literature as vertical broken
lines in Fig.~\ref{f:mabshist}.  The galaxy surveys used were chosen to
sample similar redshifts as the ELG samples and to be broadly
representative of the field galaxy population.  For the \Halpha\
emitting ELGs we show $M_B^\star$ derived from the Sloan Digital Sky
Survey luminosity function \citep[][$M_B^\star = -20.31$]{blanton03} and
the Two Degree Field Galaxy Redshift Survey \citep[][$M_B^\star =
-20.56$]{madgwick02}, while for the \fion{O}{III} and \fion{O}{II}
emitters we use the K band selected rest frame $B$ band GOODS luminosity
function results at z = 0.46 ($M_B^\star = -21.43$) and 0.97 ($M_B^\star
= -21.45$) respectively \citep{dahlen05}.  

In all cases the ELG samples have median $M_B$ values below that of the
field population at the same redshift.  As one might expect, the
difference between $M_B^\star$ and the median $M_B$ of the ELGs depends
on the redshift and hence the lines identified in our survey.  There is
an interesting difference in the histograms at the bright end.  There
are no \Halpha\ nor \fion{O}{III} emitters with $M_B < -21$, while seven
of the eleven \fion{O}{II} emitters are brighter than this value, with
the most luminous, object \#44, having $M_B = -23.5$.  This difference
may be due to sample size and relative volume: the volume available to
\fion{O}{II} emitters is 14 times that available to \Halpha\ emitters
and 2.7 times that available to \fion{O}{III} emitters. Hence if the
samples all had the same parent luminosity function, then the density of
\fion{O}{II} emitters implies that we might expect about three \fion{O}{III}
emitters with $M_B < -21$ in our survey.  A larger sample is needed to
determine if the apparent deficit of high luminosity \fion{O}{III}
emitters is real.  The surface density of the most luminous \fion{O}{II}
emitters is consistent with what we know about the luminosity function
at the redshifts we sample.  For example, \citet{dahlen05} find
galaxies with $M_B < -22$ have a surface density of 0.37 arcmin$^{-2}$
from a $K_S$ band selection in the GOODS-S field, while the three
galaxies we find yield a surface density of 0.25 arcmin$^{-2}$.

The rest frame equivalent width \EW\ is effectively a measurement of the
ionizing flux relative to the underlying continuum.  The \EW\ was
calculated from the \aXe\ spectra using
\begin{equation}
EW = \frac{\Fline}{f_\lambda(1+z)},
\end{equation}
where $f_\lambda$ is the continuum flux density, measured from the
spectra.  Caution must be used in interpreting EW values, especially at
the high end, due to background subtraction uncertainties.  For \Halpha,
the EW gives an indication of the present star formation rate compared
to the past average.  The interpretation is less clear for the
\fion{O}{III} and \fion{O}{II} \EW.  \fion{O}{III} is considered a less
reliable tracer of star formation because of its metallicity dependence,
while the underlying continuum for \fion{O}{II} emission is likely to be
dominated by A stars perhaps produced in the same event causing the
\fion{O}{II} emission.

Figure ~\ref{f:ewhist} shows the \EW\ distribution of our sample
compared to a variety of galaxy samples out to moderate redshifts.
Equivalent width statistics of our sample, split by line ID, are
compiled in Table~\ref{t:elgstats}, as are the corresponding statistics
of the comparison samples.  Figure~\ref{f:ewhist}a compares all the
lines we measure to two local samples: ELGs found with prism
spectroscopy by the KPNO International Spectroscopy Survey (KISS) red
surveys \citep{kissr1,kissr2}, and \HI\ selected ELGs imaged with
narrow-band filters for the Survey of Ionization in Neutral Gas Galaxies
\citep[SINGG][]{singg1}.  The grism-selected ELGs have significantly
higher EWs.  The differences between the samples may largely be due to
selection,  measurement or instrumentation differences.
While both the KISS and our grism survey use slitless spectroscopy, the
KISS survey has a higher dispersion (24 \AA\ pixel$^{-1}$), and employs
a filter that limits the spectral range to cover 800\AA, thus limiting
the sky background.  This makes it easier to detect lower EW systems.
Because of the higher spatial resolution of our data, the ACS spectra
have extraction aperture widths of typically 0.25\as, more than an order
of magnitude smaller than the 4\as-5\as\ resolution of the KISS survey.
This means that the continuum in the ACS grism spectra are less diluted
by non line-emitting portions of the host galaxy.  The SINGG survey uses
a totally different technique - narrow-band images to isolate \Halpha\
and $R$ band images for continuum subtraction accurate to a few
Angstroms, allowing even lower EW values to be measured.  Consequently
SINGG includes many low surface brightness and low EW systems.

Figure~\ref{f:ewhist}b-d splits our sample by line ID and compares the
EW histograms to those from the Canada-France Redshift Survey
\citep[CFRS][]{cfrs14} and \citep[SPS][]{teplitz03a,teplitz03b} for
\Halpha\ emission (panel b), \fion{O}{III} emission (panel c), and
\fion{O}{II} emission (panel d).  The CFRS is an $I$ band selected
terrestrial spectroscopic survey, while the SPS is a {\it HST\/}
slitless spectroscopic survey of effectively random high-latitude
fields.  Both comparison samples have numerous detections of the three
lines of interest and samples that extend out to $z \approx 1.5$.  Our
survey typically finds higher \EW\ values for all lines than found in
the CFRS.  Again, this could in part be due to instrumentation
differences; the CFRS spectra were obtained through 1.75\as\ wide
slitlets \citep{cfrs2}, typically covering a large fraction of the
galaxy, and about an order of magnitude larger than the extraction
apertures we use.  The SPS data is closer in nature to ours.  The STIS
slitless spectral resolution is significantly finer than our data - the
two pixel resolution element corresponds to $\sim 10$\AA; allowing STIS
to detect lower \EW\ features.  Despite that, the SPS \EW\ distributions
are broadly similar to ours.  This suggests that the relatively high EWs
seen by SPS and ourselves compared to the CFRS may result from the
smaller projected aperture sizes afforded by the space based
observations.

We next consider the broad band optical colors of ELGs in order to
assess whether they can easily be selected by color.  The \bi\ versus
\iz\ two color diagram of the ELGs identified in this study are compared
to all HDFN galaxies in Fig.~\ref{f:biiz}.  The colors were chosen
because they span the broadest $\lambda$ coverage of the WFC filter set,
and because they do a reasonable job at separating sources by redshift.
Only sources with $\zphot < 1.62$ are included to limit the comparison
sample to galaxies which would have \fion{O}{II}, or lines to the red,
at measured $\lambda \leq 9750$\AA\ where they could be detected with
the grism.  While there are very few ELGs in the red tail of the field
galaxy color distribution, the ELGs are not particularly blue.  This can
be seen in Table~\ref{t:colorstats} which tabulates the median, first
and third quartile colors of the two samples.  The samples are
subdivided by \zphot\ (field galaxies) and line identification (ELGs) so
that colors at similar redshifts are compared.  In general, ELGs do not
show as much of a ``red tail'' to their color distributions but are
otherwise similar to the general field population.  This is seen by
their third quartile colors which are distinctly bluer for the ELGs
compared to the field galaxies, while the first quartile and median
colors are within $\sim$ 0.2 mag of the field galaxies.  There are two
exceptions - the \Halpha\ emitters are somewhat redder than low-redshift
field galaxies in \bi\ and the \fion{O}{II} emitters are distinctly
bluer than the field galaxies, especially in \bi.  We attribute the
latter to the shift in $\lambda_0$ of the filters with redshift.  At $z
= 0.15, 0.53, 1.13$ the \bi\ color samples rest frame colors
$m_{378}-m_{674}$, $m_{284}-m_{507}$, $m_{204}-m_{363}$\footnote{the
  number in the subscript gives the approximate central $\lambda$ in
  nm.} - hence we are measuring rest frame ultraviolet colors for the
\fion{O}{II} emitters, while the \bi\ colors for the other line emitters
is still largely an optical color.  Presumably the ELGs are mostly
star-forming galaxies.  This suggests that the rest-frame optical colors
of ELGs are fairly normal, but the rest frame UV colors are blue.  This
is reasonable, and can be seen in the template spectra used in
photometric redshift estimates.  When normalized to unity at $\lambda =
4000$\AA\ there is relatively little variation in SED shape for $\lambda
> 4000$\AA\ but strong color variations with UV color getting bluer as
spectral type becomes later.

A better separation of the line emitters can be obtained by including
broad band filters that extend further to the blue than WFC's filter
set.  Figure~\ref{f:uvbz} shows that the $(U-V)$ versus $(B-z)$ diagram
is particularly well suited for emission-line identification.  We
examined a variety of optical two color diagrams and found this to be
the best at discriminating between line emitters.  If data in only three
filters can be obtained, then the $(U-V)$ versus $(V-z)$ diagram also
provides reasonable color separation between the emitters of the
different lines.

\section{Line identification and Redshift}\label{s:compare}

Figure~\ref{f:zgrzsp} compares the redshifts from the grism data with
the spectroscopic redshift measurements from \citet{cbhcs04}.  Symbol
shape and color are used to indicate the line identification, while
closed and open symbols mark measurements with methods A and B
respectively.  The dispersion in $\zgrism - \zspec$ about the unity line
is 0.016 for method A (15 measurements) and 0.009 for method B (20
measurements) after applying an iterative 3$\sigma$ rejection (1
measurement was rejected in each case).  We attribute the lower scatter
from method B as reflecting its superior accuracy in measuring
$\lambda$ in off-center star formation knots.

The redshift accuracy depends on how secure the line identification is.
For the relatively bright sources plotted in Fig.~\ref{f:zgrzsp}, 
the identification can be considered secure in the cases where more than
one line is identified or where the line is identified with \Halpha\ or
\fion{O}{II}.  These two lines are relatively isolated and usually there
is no other plausible line within the redshift uncertainty of the first
guess.  In these cases, the dispersion in $\zgrism - \zspec$ about the
unity line is 0.010 for method A (10 measurements) and 0.007 for method
B (13 measurements).  The line identification is not secure if there is
only one line and we identify it as \fion{O}{III}.  The line could also
be \Hbeta; however, by default \fion{O}{III} is adopted as the
identification under the assumption that typically the \fion{O}{III}
doublet is stronger than \Hbeta\ \citep[cf][]{cfrs14}.  There are 5 (6)
ELGs with spectroscopic $z$ and single lines identified as \fion{O}{III}
in our data and found with method A (B); the dispersion of their
residuals about the unity line is 0.025 (0.009)

Single-line identification depends critically on the first guess
redshift.  We find that in $\sim$ 90\%\ of the cases with both
spectroscopic and photometric redshifts that \zphot\ is sufficiently
accurate to get the correct line identification.  Specifically, when
\zspec\ is not employed as one of the redshift guesses 1 out of 16 of
the method A line identifications changes, while 2 out of 20 of the
method B identifications change.  Of course, one also has to be careful
with the spectroscopic redshifts; as noted by \citet{lfy98} the
misidentification of spectroscopic sources previously had resulted in
large discrepancies between \zspec\ and \zphot\ in the HDFN.  This is
likely to be the case for ELG \#37, which is the outlier in
Fig.~\ref{f:zgrzsp}.  It has $\zspec = 0.341$ in the Hawaii catalog
which would imply that we might see \Halpha\ at $\lambda = 8802$\AA\ or
\fion{O}{III} at $\lambda = 6657$\AA.  No features are seen near either
wavelength.  Instead we see a strong line at $\lambda = 7467$\AA.  This
source has $\zphot = 0.97$ from C04 and $\zphot = 0.92$ from our own BPZ
results.  Using either of these produces a first guess line
identification as \fion{O}{II} at $\zgrism = 1.00$, well
within the expected redshift errors of both \zphot\ estimates.

The reliability of our \zgrism\ estimates is similar to that seen for
photometric redshifts, while the accuracy is much better.  This can be
seen by comparing photometric and spectroscopic redshifts
\citep[from][]{cbhcs04} for the sources in the field.  The dispersion
about the unity line in $\zphot - \zspec$ is 0.073, 0.107, 0.082 for
\zphot\ estimates from C04, FLY99, and our BPZ results respectively.
Here we adopted a $|\zphot - \zspec| > 0.32$ rejection criterion and
only considered sources with $\zspec < 1.5$.  The rejection criterion
corresponds to three times the dispersion in $\zphot - \zspec$ from
FLY99 calculated with a 3$\sigma$ clipping, while the \zspec\ limit is
adopted to correspond to the observed \zgrism\ range of our survey.  The
number of sources rejected/used in these calculations are 11/160, 3/103,
and 12/107, respectively.  Hence $\sim$ 5\%\ to 10\%\ of photo-z
estimates are significantly discrepant compared to \zspec.  This is
similar to the reliability of our line identifications using \zphot\ as
the first guess redshifts.  More importantly, the \zgrism\ results are
more accurate than \zphot\ (have a lower dispersion in $z - \zspec$) by
a factor of 5 to 12.

The above comparisons, of course, require a spectroscopic redshift.  As
illustrated in Fig.~\ref{f:mhist} these correspond to the brighter
galaxies.  The mean $i_{775}$ is 22.7 ABmag for the 22 ELGs we selected which
also have \zspec.  The 24 ELGs without \zspec\ have a mean $i_{775} =
25.0$.  For them the main reliability issue is the photometric error
bars.  The errors are typically larger for faint galaxies, and in six of
these galaxies they translate into redshift errors large enough to allow
alternate bright line identifications.  Choice of \zphot\ source can
also be an issue.  In eight cases the line identification changes
depending on which first guess redshift is used.  In one galaxy without
\zspec, no line identifications are allowed in the range of allowed
\zphot\ from our one estimate of \zphot.  There are a total of 14 cases
that are ambiguous in one or more of these ways, 12 of those do not have
a \zspec\ estimate.  The $i_{775}$ ABmag distribution of the sources
with ambiguous line identification, are marked in gray in the top panel
of Fig.~\ref{f:mhist}, illustrating that the ambiguous identifications
correspond to faint sources (mean $i_{775} = 25.46$ ABmag).  These cases
are identified in Table~\ref{t:source} with alternate line
identifications noted in the appendix.  While some uncertainty remains
for these objects, we emphasize that by selecting the line
identification closest to the favored \zphot\ and prioritizing our
\zphot\ sources, we increase the probability that we have picked the
right line.

While one could hope that additional priors might remove the ambiguity
of the line identifications, we have not found a satisfactory
measurement to use.  For example, \citet{droz05} decide on line ID, in
part, by looking at the size of the host galaxies.  However, size alone
is not a great indicator of redshift.  This is demonstrated in
Fig.~\ref{f:sizz} which shows the size versus redshift relationship of
the objects in our field.  There is little if any angular size evolution
with redshift.  

Correlations between the luminosity and line ratios have also been
suggested to us for improving identifications.  For example, the
mass-metallicity relationship \citep{tremonti04} is in the sense that
that galaxies with low mass, and hence low luminosity, have low
metallicities, resulting in high excitations and thus typically higher
\fion{O}{III}/\Hbeta\ and \fion{O}{III}/\fion{O}{II} ratios: low
luminosity galaxies are more likely to be \fion{O}{III} emitters while
high luminosity galaxies are more likely to be \fion{O}{II} emitters.
However, on its own luminosity is unlikely to be be useful in constraining
identifications for (at least) three reasons.  (1) Using G800L tends to
select high \EW\ systems which are more likely to be high-excitation,
low metallicity systems; this should induce a bias towards \fion{O}{III}
emitters.  The calibrating sample would need to have similar selection
effects as the grism ELGs.  (2) The mass metallicity relationship is
known to evolve with redshift \citep{savaglio05}; higher redshift
galaxies of the same stellar mass have lower metallicities (again
favoring \fion{O}{III} emitters). (3) The luminosity-redshift
relationship goes in the wrong direction to remove the degeneracy.  For
example ELG \#45, one of the cases with ambiguous line identification,
has $\zgrism = 1.422$ for our adopted \fion{O}{II} identification,
yielding $M_B = -21.3$, brighter than the local $M_B^\star$ which seems
consistent with the \fion{O}{II} identification.  If we adopt the
alternative \fion{O}{III} identification then $\zgrism = 0.807$ and $M_B
= -19.2$, much fainter than $M_B^\star$, which seems to be consistent
with an \fion{O}{III} identification.  Either combination seems
plausible and the degeneracy is not broken.

One may also consider using \EW\ as a prior to decide between possible
line identifications, particularly in conjunction with luminosity.  In
the local universe it is very rare to have $EW(\fion{O}{II}\/) > 100$
\AA; less than 2\%\ of the prism selected \Halpha\ ELGs in the sample of
\citet{gzrav96} meet this condition.  Such \EW\ values are more common
from \fion{O}{III} emission.  When seen in \fion{O}{II}, the source
typically has a low (fainter than $M_B^\star$) luminosity; $M_B \gtrsim -20$
\citep[when converted to our adopted cosmology][]{pzgg00,gzrav96}.  Using
\EW\ as a prior would cast suspicion on the seven single line
\fion{O}{II} galaxies in our sample with such high EW values.  Such
scrutiny is warranted since six of these seven have some ambiguity in the
line identification (as noted in the appendix); in four of those cases an
\fion{O}{III} identification is allowed depending on which \zphot\ is
used as a first guess redshift.  If we were to use luminosity as well in
the prior, changing the \fion{O}{II} identification to \fion{O}{III} in
the ambiguous cases if \EW(\fion{O}{II}) $> 100$\AA\ and $M_B < -20.0$,
then two sources (\#37 and \#45) would be effected.

While the identifications of the high \EW\ \fion{O}{II} emitters deserve
some skepticism, at this time it would be inappropriate to apply an
$EW$ prior even in conjunction with luminosity.  There are three reasons
for this assessment. (1) As noted in Sec.~\ref{ss:stats} the continuum
levels used to determine \EW\ are prone to large background subtraction
uncertainties hence the accuracy of high \EW\ measurements are typically
low.  In this exploratory study we have not calculated the uncertainties
in the continuum level which would need to be done to properly apply a
prior.  (2) High \EW(\fion{O}{II}) values have a precedence in the more
distant universe: \citet{hogg98} find two \fion{O}{II} emitters with
$\EW > 100$\AA, while there are 22 such sources in the SPS
\citep{teplitz03a,teplitz03b} (these studies also note the problem in
continuum determination).  All these cases have $\zspec \gtrsim 0.5$.
We caution that the majority of the cases found by the SPS are also
single line sources, and thus one might also be suspicious of
their proper identification.  However, two of their high \EW\
\fion{O}{II} emitters have additional lines that secure their
identification.  (3) Finally, we note that there is strong evidence for
evolution in \EW, with $EW(\fion{O}{II})$ increasing with redshift
(especially for $z \gtrsim 0.9$), even for the most luminous galaxies
\citet{cfrs14}.  Reasons (2) and (3) indicate that an \EW\ prior based
on the local universe may not be appropriate at the redshifts we are
dealing with.  We conclude that the high \EW(\fion{O}{II}) emitters need
more scrutiny to confirm their reality.  This should include a more
careful determination of \EW(\fion{O}{II}) using an improved error
analysis, as well as follow-up spectroscopy to detect additional lines
and confirm the line identification.  If the high incidence rate of
sources having \EW(\fion{O}{II}) $> 100$\AA\ is confirmed, it would be
further proof of strong redshift evolution in the star forming
properties of galaxies.

\section{Summary and discussion}\label{s:disc}

We have shown that a modest expenditure, three orbits, of ACS WFC grism
time with the {\it HST\/} pointed at a ``blank'' high latitude field
results in the detection of dozens of ELGs out to $z \sim 1.5$.  Here we
found 46 ELGs in the HDFN yielding a surface density of 3.9 ELGs per
square arcmin.  The ``blind'' grism selection technique (method B)
results in significantly more sources and better redshift accuracy.  We
attribute this to its ability to isolate individual emission-line knots
within galaxies.  The \aXe\ selection technique (method A) relies on an
initial catalog of sources, and hence is effectively a broad band (i.e.\
usually continuum) selection technique.  While it often misses objects
where emission arises from a knot, it is adept at picking up line
emission confined to a compact nucleus, which the ``blind'' technique
can miss.  Hence, the two techniques are complimentary.

The ELGs found are most frequently \fion{O}{III} (or \Hbeta) emitters at
$z \sim 0.4$ to 0.9.  \Halpha\ and \fion{O}{II} emitters are also found,
but are less common because of the smaller volume for the former, and
the limited depth of the observations for the latter.  The ELGs
represent a small fraction of the field population.  There are 647
galaxies within the field of our observations having $\zphot \leq 1.5$
in the \citet{cap04} catalog, while 186 galaxies have spectroscopic $z
\leq 1.5$ in the \citet{cbhcs04} compilation.  While grism selection of
ELGs does not result in a sample of the field that is in any way
complete, the galaxies selected do have interesting properties.  In
particular they tend to be low-luminosity high-EW systems.  This
suggests that they are experiencing an intense burst of star formation,
or may contain an AGN.  A high \fion{O}{III} EW suggests high excitation
and low metallicity.  Grism selection of ELGs may be a good means to
locate the barely evolved building blocks of larger galaxies.

Our results are consistent with deeper G800L observations reported by
the GRAPES team \citep{grapes1,egrapes06a} who obtained 92 ks of G800L
observations of the HUDF (13 times longer than our HDF observations)
split into five epochs.  They found 113 ELGs in a field having similar
area, using an algorithm
equivalent to our method A (although differing in some details); 51 of
these are brighter than our empirical line flux limit $\Fline =
3.0\times 10^{-17}\, {\rm erg\, cm^{-2}\, s^{-1}}$. This compares well
with the 46 ELGs we find in the HDF.  They also find \fion{O}{III}
sources to be the most frequently detected line, while the maximum
\Fline\ is $2.2\times 10^{-17}\, {\rm erg\, cm^{-2}\, s^{-1}}$ in the
seven \Lya\ emitters they find, consistent with our non-detection of
these sources.

Optimal use of grism data to discover ELGs requires additional data.
This is because the grism spectra typically show only one emission line
per object, hence identification of the line from the low-resolution
spectra is difficult, at least for the relatively short exposures used
here.  With longer exposures, often both the \fion{O}{III} doublet and
\Hbeta\ lines can be seen in ELGs having $z \approx 0.4$ to 0.9, hence
the problem then becomes distinguishing between \Halpha\ and
\fion{O}{II} emitters.  

The accuracy of the line identification can be improved if there is a
good ``first guess'' redshift for each source, either a spectroscopic
redshift \zspec\ or a photometric redshift \zphot.  While the former
produces the most accurate line identifications, there is typically
little need for a grism spectrum of sources that already have a
ground-based spectrum of sufficient $S/N$ to determine a redshift.  Use
of \zphot\ as the first guess requires additional photometric data from
{\it HST\/} or other sources to derive the redshift.  Without these
additional data, or follow-up spectroscopy, it may be impossible to
identify the line and hence determine the redshift, which seriously
diminishes the utility of the ELG discoveries.  With a good \zphot\ 
first guess, lines can be identified with $\sim 90$\%\ reliability,
similar to the \zphot\ reliability, but resulting in redshifts accurate
to $\sim 0.01$ (3000 \kms).  This is about an order of magnitude better
than \zphot\ estimates and is sufficient for separating ELG members of
rich clusters from the field.

The requirement of additional photometry to obtain good emission-line
redshifts amounts to a substantial additional investment of time and
labor.  In Sec~\ref{ss:stats} we examined color-color diagrams that are
useful for sorting the ELGs by line identification.  In general filter
combinations that span the full optical range seem to be the best suited
for this purpose.  The $U - V$ versus $B - z$ diagram provides the best
discrimination, but requires wide field $U$ band data which is hard to
obtain from the ground, and impossible to obtain with ACS + WFC.  The
\bi\ versus \iz\ diagram does a reasonable job at separating the
\fion{O}{II} emitters from the \fion{O}{III} and \Halpha\ sources.  Most
of the discrimination comes from the \bi\ color which is the single
color best suited for line discrimination from the ACS + WFC filter set.
However, as shown in Fig.~\ref{f:biiz} it does not discriminate well
between \Halpha\ and \fion{O}{III} emitters.  For that one needs to have
a filter as far as possible towards short wavelengths so as to sample
the rest frame UV at modest redshifts.  An efficient solution of this
issue at {\it HST\/} resolution would require the installation of WFC3.
Until then, terrestrial $(U-V)$ gives the best single color
discriminator between line identifications.

The reliability of the line identification decreases with decreasing
brightness, as the increasing photometric errors can result in ambiguous
line identification.  The investment of direct imaging time required to
beat down the photometric redshift errors is larger than the time spent
on the grism imaging; we spent three {\it HST\/} orbits imaging with
G800L, while the GOODS direct images ($B_{435}$ - 3 orbits, $V_{606}$ -
2.5 orbits, $i_{775}$ - 2.5 orbits, $z_{850}$ - 5 orbits, at each pointing), as well as
the original HDF images and ground-based imaging were used to determine
the first guess photometric redshifts.  Even then, more than half of the
ELGS we detected with $i_{775} > 24.5$ ABmag have some uncertainty in
their line identification.  Of particular concern are the faint
\fion{O}{II} identifications that have apparently large $EW > 100$\AA.
These could signify strong redshift evolution in the star forming
population, as indicated by other studies \citep[e.g.][]{cfrs14}, or
could be (in part) spurious due to contamination of misidentified
\fion{O}{III} emitters, or large continuum placement errors. The HDFN is
one of the best studied deep field, yet we still face these issues
because of the faintness of the ELGs.

We conclude that ACS G800L grism data with minimal direct images
provides a useful means of locating ELGs, but without additional data,
can provide only a limited interpretation of the nature of the sources.  Photometric
redshifts from broad band imaging can improve the reliability of line
identifications.  However, these data are also expensive to obtain, and
the results are still likely to be ambiguous for the faintest ELGs we can detect.
Secure redshifts for these still require ground-based spectroscopy.
Fortunately the line fluxes are easily within the reach of the current
generation of 8m class telescopes.  For example the Gemini Multi-Object
Spectrographs can detect an emission line with a flux of $1.5 \times
10^{-18}$ erg cm$^{-2}$ s$^{-1}$ (one tenth of our limiting line flux)
and EW of 10\AA\ at $S/N \sim 5$, in a single 900s exposure.  An
efficient strategy for finding and characterizing ELGs would then be to
observe with ACS and the G800L grism to find the emission-line sources,
employing broad band images (in say F814W and F606W) of a similar depth
to locate the corresponding galaxy; then following up with ground based
spectroscopy to secure the redshift and identify additional lines.

\acknowledgments

ACS was developed under NASA contract NAS 5-32865, and this research has
been supported by NASA grant NAG5-7697 and by an equipment grant from
Sun Microsystems, Inc.  The {Space Telescope Science Institute} is
operated by AURA Inc., under NASA contract NAS5-26555.  We are grateful
to K.~Anderson, J.~McCann, S.~Busching, A.~Framarini, S.~Barkhouser, and
T.~Allen for their invaluable contributions to the ACS project at JHU.
GRM acknowledges useful conversations with Anna Pasquali, Marco
Sirianni, James Rhoads, Sangeeta Malhotra, Chun Xu, and S{\o}ren
Larsen.  We thank the anonymous referee for suggestions that improved
the science and readability of this paper.

\appendix

\section{Comments on individual sources\label{ss:sources}}

Here we present notes on ELGs with multiple ELSs (lines and/or knots),
those with striking morphologies, and cases where the line
identification is in some way ambiguous.  For each source we list the
ELG id number, the corresponding long name, and grism-redshift (in
parenthesis).


{\bf \#1 - GOODSN J123641.63+621132.1} (0.098): The lowest redshift
galaxy shows \Halpha\ emission from a bright knot, pinpointed with
method B, 1.07\arcsec\ to E of the spiral galaxy core.  This knot
combined with a fainter knot to the east of the core results in a weak
\Halpha\ detection with method A.

{\bf \#2 - GOODSN J123644.75+621157.4} (0.124): The adopted
identification of the single line in this source as \Halpha\ yields a
redshift closest to the C04 \zphot.  However, the error bars in C04 and
our \zphot\ also allow an identification of \fion{O}{III}, while the
\zphot\ from FLY99 only allows \fion{O}{III} as the identification.

{\bf \#3 - GOODSN J123633.16+621344.0} (0.126): The adopted \Halpha\ line
identification yields \zgrism\ closest to \zphot\ from C04.  However, the
error bars from C04 also allow line identifications of \fion{O}{III}, or
\fion{O}{II}, while our BPZ results allow \Halpha\ and \fion{O}{III} as
the line identification.

{\bf \#5 - GOODSN J123648.30+621426.9} (0.136): Lopsided spiral, with a
bright compact nucleus and bar.  The spectrum extracted with method A
shows weak \Halpha\ on top of a strong continuum.  With method B, four
\Halpha\ emitting knots are identified with lines at a consistent
$\lambda$ (average $\lambda = 7456$\AA).  A fifth ELS appears to have
emission at a discrepant $\lambda$ perhaps due to misidentification of
the emitting knot in the direct image.

{\bf \#7 - GOODSN J123658.06+621300.8} (0.308): Inclined disk galaxy
with a prominent knot offset by 0.35\arcsec.  \fion{O}{III} and \Hbeta\ 
emission arise from the galaxy nucleus.

{\bf \#8 - GOODSN J123626.57+621321.2} (0.318): Our BPZ analysis
provides the only \zphot\ source.  The \zphot\ error bars allow the
single line to be identified as \Halpha, \fion{O}{III}, or \fion{O}{II}.
The adopted \fion{O}{III} identification is closest to the third
strongest \zphot\ probability peak from BPZ.

{\bf \#9 - GOODSN J123650.82+621256.3} (0.319): Edge-on disk galaxy with
two emission-line knots separated by 0.96\arcsec\ bracketing the
nucleus.  With method B, \fion{O}{III} is clearly visible in both knots,
\Halpha\ is clearly detected in the southern knot, but is just below the
detection limit in the northern knot.  No emission lines are detected
with method A, because the extracted spectrum does not fully contain the
knots.

{\bf \#10 - GOODSN J123646.59+621157.5} (0.341): The error bars on the C04
\zphot\ allow \fion{O}{III} or \fion{O}{II} as the line identification
for the single detected line.  The closest \zgrism\ match is with \fion{O}{III}
which is also consistent with the FLY99 and BPZ \zphot\ analysis.

{\bf \#14 - GOODSN J123637.56+621240.4} (0.445): This galaxy is apparently
interacting with GOODSN J123637.64+621241.3, which we also detect as an 
ELG(see below).  We find one line, \fion{O}{III}, with method B.

{\bf \#15 - GOODSN J123637.64+621241.3} (0.446): The dominant system in the
pair with \#14.  Method B detects two lines
identified as \fion{O}{III} and \Halpha, while only one low EW line,
\fion{O}{III}, is identified with method A. 

{\bf \#17 - GOODSN J123657.30+621300.0} (0.465): A barred spiral with
arms forming a pseudo ring.  Line emission originates in an arm \HII\
region about 1\arcsec\ from the galaxy center.

{\bf \#20 - ACS J123636.58+621336.8} (0.478): This source is not present in the
GOODSN r1.1z catalog, probably due to its faintness.  Our own
measurements of this source from data combining the GOODS images and
other ACS images of the field yield photometry [$B_{435}$, $V_{606}$,
$i_{775}$, $z_{850}$\/] = [$28.16\pm 0.14$, $27.69\pm 0.08$, $26.77\pm
0.06$, $27.72\pm 0.15$] ABmag, while measurements from the \citet{cap04}
images yield [$U$, $B$, $V$, $R$, $I$] = [$29.16\pm 0.55$,
$28.22\pm0.30$, $28.00\pm 0.28$, $28.11\pm 0.34$, $27.7\pm 0.6$] ABmag
and $z > 26$ ABmag (a non-detection).  The relative brightness in
$i_{775}$ is likely due to the single bright line we observe at $\lambda
= 7386$\AA\ whose flux (Table~\ref{t:source}) is consistent with
dominating $i_{775}$.  Using this photometry, BPZ yields a best $\zphot
= 0.62$ consistent with the line being \fion{O}{III}.  However a second
peak in the probability distribution at $\zphot = 0.1$ means that an
identification as \Halpha\ can not be ruled out.  Because this source
was not in the GOODSN r1.1z catalog, it was excluded from the statistics
given in Table~\ref{t:colorstats} (below).

{\bf \#22 - GOODSN J123655.58+621400.3} (0.551): Emission corresponds to a knot
above the plane of an edge-on disk galaxy.

{\bf \#24 - ACS J123657.48+621212.0} (0.555): At first blush, this object
appears to be the nucleus of a dwarf galaxy being shredded by an
interaction with its neighbor GOODSN J123657.49+621211.2 projected
1.46\arcsec\ towards SSE.  However, that source has $\zspec = 0.669$
(C04).  Our redshift is from a single line detected with both methods
identified as \fion{O}{III}.  Even if the line were \Hbeta\ at $z =
0.600$, or \Hgamma\ at $z = 0.790$, the redshift would be significantly
discrepant with its neighbor.  Hence, the apparent interaction may be
spurious and the sources an unrelated chance projection.

{\bf \#26 - GOODSN J123654.39+621434.7} (0.573): The nucleus of this
modestly inclined disk galaxy shows \fion{O}{III} and \Hbeta.  The
emission lines are found with method A, not B, probably because the line
emission is centered on the compact nucleus.

{\bf \#27 - GOODSN J123645.53+621330.2} (0.670): The large error bars on
the C04 \zphot\ allow the single line to be identified as \Halpha,
\fion{O}{III} or \fion{O}{II}; the closest \zgrism\ match is the adopted
\fion{O}{III} identification.  This is consistent with the smaller error
bars on the \zphot\ from BPZ, while none of the brightest likely lines
match the \zphot\ from FLY99.

{\bf \#28 - GOODSN J123636.47+621419.1} (0.684): The off-center knot in
this small galaxy shows a broad, bright emission line well fit as
\fion{O}{III} blended with \Hbeta.

{\bf \#30 - GOODSN J123647.24+621134.7} (0.717): BPZ is the only source of
\zphot\ for this faint source, and yields allowed redshifts in the range
$1 \lesssim \zphot \lesssim 2$.  However, none of the typical bright lines
can match this range and the observed $\lambda$.  The adopted
\fion{O}{II} identification corresponds to the line closest to the
allowed range. 

{\bf \#33 - GOODSN J123644.17+621430.5} (0.858): The large C04 \zphot\ 
error bars for this source allows the single line to be identified as
\fion{O}{II} as well as the preferred \fion{O}{III}.  The \zphot\ from
BPZ is consistent with our adopted \fion{O}{III} line identification.

{\bf \#34 - GOODSN J123652.97+621257.1} (0.943): The preferred \zphot\
from FLY99 identifies the single line as \fion{O}{II}, while the BPZ
\zphot\ indicates the line may be \fion{O}{III}.

{\bf \#35 - GOODSN J123636.63+621347.1} (0.947) This bright compact
galaxy is detected in both \Hbeta\ and \Hgamma.  It is the only \Hgamma\
source in the sample.  The flux ratio $F_{\rm H\beta}/F_{\rm H\gamma}
\sim 1.9$ is close to the expected case B ratio of 2.1 \citep[for $n_e =
100$ cm$^{-2}$ and $T_e = 10^4$ K][]{ds03}.  The lines are detected with
method A but not with method B, probably because the line emission is
centered on the compact nucleus.  The case for interaction with GOODSN
J123636.85+621346.2, a larger but slightly fainter lopsided spiral
1.81\arcsec\ to ESE, is strong since one of its spiral arm seems to be
connected to GOODSN J123636.63+621347.1 in the high-pass filtered direct
image, reminiscent of the M51/NGC5194 system (see Fig~\ref{f:stamps}).
However, the spectroscopic redshift of the spiral is significantly lower
(0.846, C04) casting some on this inference.  The \zphot\ from C04 are
consistent with these line identifications, while our BPZ \zphot\ is too
low.

{\bf \#36 - GOODSN J123649.35+621155.4} (0.954): Compact galaxy with two
lines arising in the nucleus identified as \fion{O}{II} and
\fion{Ne}{III}$\lambda$3869\AA.  The measured line ratio $F_{[NeIII]}/F_{[OII]} \sim
0.38$ indicates a high excitation: it corresponds to the 80th percentile
in this ratio for the local ELGs which display both lines in the catalog
of \citet{tmmmc91}.  

{\bf \#37 - GOODSN J123649.47+621456.9} (1.003): This source has the largest
discrepancy between \zspec\ and \zgrism\ in Fig.~\ref{f:zgrzsp}.  The
original source for the reported $\zspec = 0.341$ is \citet{c00}, where
the spectrum is given a quality code of 5: ``one emission line only,
reality uncertain, assume 3727Å'' \citep{c99}, that is, a single weak
line in the spectrum.  Both our BPZ results and those of C04 indicate
very similar \zphot: 0.921, and 0.970, respectively.  While there is no
good match at $z \approx 0.34$ for the line we detect at
$\lambda = 7469$\AA, adopting either \zphot\ estimate for our first
guess redshift identifies the line as \fion{O}{II} at our adopted
$\zgrism = 1.003$.  Using this redshift, then the line found by
\citet{c00} may be \fion{O}{II}2470\AA\ or \fion{Ne}{IV}2423\AA, if the
weak line they found is real.

{\bf \#40 - GOODSN J123645.46+621357.3} (1.073): The preferred \zphot\ from
FLY99 yields the single line identification of \fion{O}{II},
while the BPZ \zphot\ indicates the line may be \fion{O}{III}.

{\bf \#42 - GOODSN J123653.51+621141.4} (1.263): The C04 \zphot\ yields
our adopted \fion{O}{II} line identification.  The \zphot\ from BPZ is
lower and outside the C04 error-bars, but does not allow any of the
standard bright line guesses to correspond with the observed line
$\lambda$.

{\bf \#44 - GOODSN J123652.77+621354.7} (1.346) The brightest knot of this
chain galaxy has one line identified as \fion{O}{II}.  

{\bf \#45 - GOODSN J123642.55+621150.3} (1.422): The adopted \fion{O}{II}
identification for the single line of this source is consistent with
\zphot\ from C04 and FLY99, while BPZ prefers a lower \zphot\ consistent
with an \fion{O}{III} identification.

{\bf \#46 - GOODSN J123648.48+621120.7} (1.424): BPZ is the only \zphot\
source for this faint source; the error bars allow an \fion{O}{III}
identification as well as the adopted \fion{O}{II}.


\begin{deluxetable*}{l c c c}
  \tablewidth{0pt}
  \tablecaption{HDFN observations\label{t:obs}}
  \tablehead{\colhead{filter} &
             \colhead{exp.\ time} &
             \colhead{$N_{\rm exp}$} &
             \colhead{Resolution [pix]}}
\startdata
G800L  & 6870 & 6 & 2.14 \\
F775W  & 4500 & 4 & 1.72 \\
F850LP & 6800 & 6 & 1.83 \\
\enddata
\end{deluxetable*}

\begin{deluxetable*}{r r c c c c c c c c l l }
  \tabletypesize{\footnotesize}
  \tablewidth{0pt}
  \tablecaption{Emission line sources\label{t:source}}
  \tablehead{\colhead{ELG \#} &
             \colhead{Name} &
             \colhead{$i_{775}$} &
             \colhead{\zspec} &
             \colhead{\zphot} &
             \colhead{S} &
             \colhead{\zgrism} &
             \colhead{ID} &
             \colhead{EW} &
             \colhead{$\log(F_{\rm line})$} &
             \colhead{M} &
             \colhead{Notes} \\
             \colhead{(1)} &
             \colhead{(2)} &
             \colhead{(3)} &
             \colhead{(4)} &
             \colhead{(5)} &
             \colhead{(6)} &
             \colhead{(7)} &
             \colhead{(8)} &
             \colhead{(9)} &
             \colhead{(10)} &
             \colhead{(11)} &
             \colhead{(11)}}
\startdata
   1 & GOODSN J123641.63+621132.1 & 19.78 & 0.089   & 0.090   & C       & 0.098   & \Halpha       &   127& $-15.52$ & BA & b       \\
   2 & GOODSN J123644.75+621157.4 & 25.06 & \nodata & 0.210   & C       & 0.124   & \Halpha       &   223& $-16.43$ & B  & cd      \\
   3 & GOODSN J123633.16+621344.0 & 25.31 & \nodata & 0.220   & C       & 0.126   & \Halpha       &   155& $-16.59$ & BA & d       \\
   4 & GOODSN J123646.53+621407.9 & 23.95 & 0.130   & 0.160   & C       & 0.128   & \Halpha       &   115& $-16.33$ & BA &         \\
   5 & GOODSN J123648.30+621426.9 & 19.01 & 0.139   & 0.130   & C       & 0.136   & \Halpha       &    62& $-15.52$ & BA & b       \\
   6 & GOODSN J123651.72+621220.5 & 21.62 & 0.300   & 0.320   & C       & 0.302   & \Halpha       &    58& $-16.20$ & BA &         \\
   7 & GOODSN J123658.06+621300.8 & 22.45 & 0.319   & 0.310   & C       & 0.308   & \Hbeta        &     9& $-16.68$ & B  & a       \\
     &                            &       &         &         &         &         & [OIII]        &     8& $-16.78$ & B  & a       \\
   8 & GOODSN J123626.57+621321.2 & 26.56 & \nodata & 0.483   & B       & 0.318   & [OIII]        &  6418& $-16.68$ & A  & d       \\
   9 & GOODSN J123650.82+621256.3 & 22.61 & 0.319   & 0.310   & C       & 0.319   & [OIII]        &   250& $-16.00$ & B  & ab      \\
     &                            &       &         &         &         &         & \Halpha       &   169& $-16.30$ & B  & a       \\
  10 & GOODSN J123646.59+621157.5 & 25.77 & \nodata & 0.440   & C       & 0.341   & [OIII]        &   248& $-16.44$ & BA & d       \\
  11 & GOODSN J123653.10+621438.4 & 24.53 & \nodata & 0.430   & C       & 0.376   & [OIII]        &   123& $-16.49$ & B  &         \\
  12 & GOODSN J123628.76+621335.8 & 25.99 & \nodata & 0.559   & B       & 0.427   & [OIII]        &   204& $-16.56$ & BA &         \\
  13 & ACS    J123632.69+621239.1 & 23.15 & 0.458   & 0.420   & C       & 0.433   & [OIII]        &    65& $-16.46$ & B  &         \\
  14 & GOODSN J123637.56+621240.4 & 22.23 & 0.457   & 0.153   & B       & 0.445   & [OIII]        &    12& $-16.58$ & B  & c       \\
  15 & GOODSN J123637.64+621241.3 & 21.08 & \nodata & 0.436   & B       & 0.446   & [OIII]        &    81& $-16.09$ & BA & a       \\
     &                            &       &         &         &         &         & \Halpha       &   155& $-15.99$ & B  & a       \\
  16 & GOODSN J123650.79+621221.7 & 24.71 & \nodata & 0.440   & C       & 0.450   & [OIII]        &   254& $-16.27$ & BA &         \\
  17 & GOODSN J123657.30+621300.0 & 21.40 & 0.473   & 0.420   & C       & 0.465   & [OIII]        &    89& $-16.41$ & B  &         \\
  18 & GOODSN J123644.19+621248.2 & 21.62 & 0.555   & 0.540   & C       & 0.476   & [OIII]        &    20& $-16.05$ & A  &         \\
  19 & GOODSN J123637.76+621235.6 & 23.86 & \nodata & 0.480   & C       & 0.477   & [OIII]        &   172& $-16.36$ & B  &         \\
  20 & ACS    J123636.58+621336.8 & 26.77 & \nodata & 0.620   & C       & 0.478   & [OIII]        &   315& $-16.45$ & BA & dg      \\
  21 & GOODSN J123645.24+621108.9 & 23.41 & 0.513   & 0.580   & C       & 0.505   & [OIII]        &   162& $-16.14$ & BA &         \\
  22 & GOODSN J123655.58+621400.3 & 24.17 & 0.559   & 0.590   & C       & 0.551   & [OIII]        &   147& $-16.44$ & BA &         \\
  23 & GOODSN J123645.32+621143.2 & 23.85 & 0.557   & 0.570   & C       & 0.552   & [OIII]        &   198& $-16.15$ & BA &         \\
  24 & ACS    J123657.48+621212.0 & 23.29 & \nodata & 0.720   & F       & 0.555   & [OIII]        &    26& $-16.59$ & BA &         \\
  25 & GOODSN J123644.75+621144.1 & 24.87 & \nodata & 0.670   & C       & 0.562   & [OIII]        &    91& $-16.54$ & B  &         \\
  26 & GOODSN J123654.39+621434.7 & 22.26 & 0.577   & 0.690   & C       & 0.573   & [OIII]        &    13& $-16.50$ & A  & a       \\
     &                            &       &         &         &         &         & \Hbeta        &    11& $-16.54$ & A  & a       \\
  27 & GOODSN J123645.53+621330.2 & 25.38 & \nodata & 0.540   & C       & 0.670   & [OIII]        &   243& $-16.47$ & BA & cd      \\
  28 & GOODSN J123636.47+621419.1 & 24.30 & \nodata & 0.700   & C       & 0.684   & [OIII]        &   341& $-15.81$ & BA & a       \\
     &                            &       &         &         &         &         & \Hbeta        &   115& $-16.47$ & B  & a       \\
  29 & GOODSN J123646.96+621133.0 & 24.27 & \nodata & 0.700   & C       & 0.685   & [OIII]        &   198& $-16.12$ & BA &         \\
  30 & GOODSN J123647.24+621134.7 & 27.10 & \nodata & 1.518   & B       & 0.717   & [OII]         &   121& $-16.24$ & A  & f       \\
  31 & GOODSN J123629.72+621329.9 & 22.83 & 0.746   & 0.700   & C       & 0.737   & [OIII]        &    28& $-16.15$ & A  &         \\
  32 & GOODSN J123642.29+621429.9 & 23.77 & \nodata & 0.850   & C       & 0.841   & [OIII]        &   287& $-16.04$ & B  &         \\
  33 & GOODSN J123644.17+621430.5 & 24.66 & 0.863   & 0.925   & B       & 0.858   & [OIII]        &   175& $-16.23$ & BA & cd      \\
  34 & GOODSN J123652.97+621257.1 & 25.63 & \nodata & 0.800   & F       & 0.943   & [OII]         &   182& $-16.32$ & BA & c       \\
  35 & GOODSN J123636.63+621347.1 & 21.44 & 0.962   & 0.386   & B       & 0.947   & \Hgamma       &     7& $-16.34$ & A  & a       \\
     &                            &       &         &         &         &         & \Hbeta        &    14& $-16.07$ & A  & a       \\
  36 & GOODSN J123649.35+621155.4 & 23.41 & 0.961   & 1.115   & B       & 0.954   & [OII]         &    32& $-16.39$ & BA & a       \\
     &                            &       &         &         &         &         & [NeIII]       &    13& $-16.81$ & B  & a       \\
  37 & GOODSN J123649.47+621456.9 & 24.11 & 0.341   & 0.970   & C       & 1.003   & [OII]         &   107& $-16.69$ & B  & ce      \\
  38 & GOODSN J123654.45+621152.8 & 24.40 & \nodata & 1.040   & C       & 1.017   & [OII]         &    36& $-16.63$ & B  &         \\
  39 & GOODSN J123658.30+621214.5 & 23.33 & 1.020   & 0.970   & C       & 1.026   & [OII]         &    40& $-16.32$ & B  &         \\
  40 & GOODSN J123645.46+621357.3 & 26.17 & \nodata & 0.920   & F       & 1.073   & [OII]         &   425& $-16.37$ & BA & c       \\
  41 & GOODSN J123643.42+621151.9 & 23.12 & 1.241   & 1.200   & C       & 1.237   & [OII]         &    22& $-16.55$ & BA &         \\
  42 & GOODSN J123653.51+621141.4 & 23.89 & \nodata & 1.490   & C       & 1.263   & [OII]         &    46& $-16.35$ & BA & c       \\
  43 & GOODSN J123644.98+621240.0 & 24.07 & \nodata & 1.170   & C       & 1.337   & [OII]         &   578& $-16.20$ & A  &         \\
  44 & GOODSN J123652.77+621354.7 & 22.73 & 1.355   & 1.440   & F       & 1.346   & [OII]         &    41& $-16.20$ & B  &         \\
  45 & GOODSN J123642.55+621150.3 & 24.97 & \nodata & 1.600   & C       & 1.422   & [OII]         &   228& $-16.57$ & B  & c       \\
  46 & GOODSN J123648.48+621120.7 & 26.87 & \nodata & 1.141   & B       & 1.424   & [OII]         &   198& $-16.49$ & B  & d       \\
\enddata
\tablecomments{Column descriptions: (1) ELG catalog number (this work). (2) Names preceded by
  GOODSN are the IAU specified name from the GOODS release r1.1z
  (Giavalisco 2004a,b).  Names preceded by ACS could not be matched with
  the GOODSN catalog.   (3) $i_{775}$ are SExtractor magnitudes of our
  images through the F775W filter in the ABmag system. (4) The
  spectroscopic redshift \zspec\ taken from the Hawaii group compilation
  (Cowie \etal\ 2004). (5) and (6) are the adopted photometric redshift \zphot\ and
  its source: C - Capak (2004); F - Fernandez-Soto et al.\ (2004); and B
  - our own BPZ measurements using GOODS photometry.  
  Entries in these two columns correspond to cases that do not match the
  (7) Adopted grism redshift, \zgrism.  (8) Adopted line 
  identification.  The one case where \zspec\ and \zphot\ are blank do not have
  good first redshift guesses, hence column (8) is left blank and the
  measured emission line wavelength, $\lambda$ in \AA\ is given in (7).  (9) Rest frame 
  equivalent width in \AA.  For the one case where \zgrism\ is undefined, the 
  observed EW is listed.  (10)
  The logarithm of the measured line flux in erg cm$^{-2}$ s$^{-1}$.
  (11) Gives the methods that detected the line emission: A - aXe
  selection, B - blind grism selection.  When a line is identified by
  both methods then the data from columns (7) - (9) are taken from
  method B.  (12) Notes: a - two lines identified; b - 
  multiple line-emitting knots; c - \zgrism\ differ depending on \zphot\ 
  source; d - large \zphot\ errors result in ambiguous line ID; e - \zspec\ 
  and \zgrism\ do not agree; f - no line identification allowed within 
  \zphot\ error bars, nearest expected line chosen; g - photometry not from  
  GOODS release r1.1z, see individual object notes.  Blank entries 
  for columns (1)-(7) and (21) occur for the second emission line in sources with two 
  emission lines (note a).  These blank entries thus have the same values as for the 
  previous line}
\end{deluxetable*}

\begin{deluxetable*}{l c c c}
  \tabletypesize{\footnotesize}
  \tablewidth{0pt}
  \tablecaption{Redshift, absolute magnitude, and $EW$ statistics of
  ELGs split by line identification\label{t:elgstats}}
  \tablehead{
    \colhead{Property} & 
    \colhead{\Halpha} &
    \colhead{\fion{O}{III}} & 
    \colhead{\fion{O}{II}}  }
\startdata
\multicolumn{4}{c}{Grism-selected ELGs (this study)}\\
Number              &     8 & 23/26\tablenotemark{a} &    13 \\
Redshift, $z$ & & & \\
~~ minimum          & 0.098 & 0.308 & 0.717 \\
~~ 25th percentile  & 0.126 & 0.436 & 1.003 \\
~~ median           & 0.132 & 0.478 & 1.073 \\
~~ 75th percentile  & 0.306 & 0.570 & 1.337 \\
~~ maximum          & 0.446 & 0.858 & 1.424 \\
$B$ band absolute mag, $M_B$ [ABmag]\\
~~ minimum          & $-20.92$ & $-20.97$ & $-23.46$ \\
~~ 25th percentile  & $-18.97$ & $-20.19$ & $-21.87$ \\
~~ median           & $-18.29$ & $-18.93$ & $-21.27$ \\
~~ 75th percentile  & $-14.17$ & $-18.00$ & $-19.93$ \\
~~ maximum          & $-13.03$ & $-14.32$ & $-16.61$ \\
rest frame equivalent width, $EW$ [\AA] \\
~~ minimum          &    58 &     8 &    22 \\
~~ 25th percentile  &   102 &    69 &    40 \\
~~ median           &   140 &   167 &   107 \\
~~ 75th percentile  &   159 &   247 &   198 \\
~~ maximum          &   223 &  6418 &   578 \\
\multicolumn{4}{c}{CFRS (Hammer et al.\ 1997)}\\
Number              &    95 &  175  & 270 \\
rest frame equivalent width, $EW$ [\AA] \\
~~ minimum          &     4 &   0.7 &   1.3 \\
~~ 25th percentile  &    27 &     7 &    16 \\
~~ median           &    41 &    15 &    27 \\
~~ 75th percentile  &    63 &    28 &    39 \\
~~ maximum          &  1520 &  1022 &   981 \\
\multicolumn{4}{c}{SPS (Teplitz et al.\ 2003a)}\\
Number              &    18 &    33 &    78 \\
rest frame equivalent width, $EW$ [\AA] \\
~~ minimum          &    19 &    13 &     6 \\
~~ 25th percentile  &    75 &    55 &    45 \\
~~ median           &   103 &   124 &    68 \\
~~ 75th percentile  &   148 &   264 &   117 \\
~~ maximum          &   394 &  1479 &   750 \\
\enddata
\tablenotetext{a}{There are 26 \fion{O}{III} emitters, however three are
missing GOODS photometry (see Table~\ref{t:source}).  Therefore we use the 23
sources with GOODS photometry to compile the $M_B$ statistics, while all
26 sources are used to compile redshift and \EW\ statistics.}
\end{deluxetable*}

\begin{deluxetable*}{l c c c r c l c c r r c l }
  \tabletypesize{\footnotesize}
  \tablewidth{0pt}
  \tablecaption{Color properties of HDFN field galaxies and ELGs\label{t:colorstats}}

  \tablehead{ & \multicolumn{6}{c}{field galaxies} & \multicolumn{6}{c}{Emission-line Galaxies} \\
             \colhead{Color} &
             \colhead{} &
             \colhead{\zphot\ range} &
             \colhead{$N$} &
             \colhead{25th} &
             \colhead{median} &
             \colhead{75th} &
             \colhead{} &
             \colhead{Lines} &
             \colhead{$N$} &
             \colhead{25th} &
             \colhead{median} &
             \colhead{75th} \\
             \colhead{(1)} & \colhead{} &
             \colhead{(2)} &
             \colhead{(3)} &
             \colhead{(4)} &
             \colhead{(5)} &
             \colhead{(6)} & \colhead{} &
             \colhead{(7)} &
             \colhead{(8)} &
             \colhead{(9)} &
             \colhead{(10)}&
             \colhead{(11)}}
  \startdata
  \bi & & $< 1.62$      & 640 &   0.63 & 1.04 & 1.40 & & all            &  45 &   0.52 & 0.87 & 1.06 \\
  \bi & & $< 0.30$      & 145 &   0.44 & 0.70 & 1.04 & & \Halpha\       &   8 &   0.79 & 0.93 & 1.11 \\
  \bi & & $0.30 - 0.76$ & 243 &   1.07 & 1.27 & 1.67 & & \fion{O}{III},\Hbeta  &  28 &   0.91 & 1.03 & 1.14 \\
  \bi & & $0.76 - 1.62$ & 261 &   0.53 & 0.87 & 1.31 & & \fion{O}{II}   &  13 &   0.39 & 0.46 & 0.73 \\ 
  \iz & & $< 1.62$      & 640 &   0.07 & 0.21 & 0.37 & & all            &  45 &   0.00 & 0.13 & 0.24 \\
  \iz & & $< 0.30$      & 145 & --0.03 & 0.06 & 0.16 & & \Halpha\       &   8 & --0.02 & 0.05 & 0.17 \\
  \iz & & $0.30 - 0.76$ & 243 &   0.06 & 0.17 & 0.27 & & \fion{O}{III},\Hbeta  &  28 &   0.00 & 0.13 & 0.20 \\
  \iz & & $0.76 - 1.62$ & 261 &   0.24 & 0.35 & 0.51 & & \fion{O}{II}   &  13 &   0.13 & 0.23 & 0.37 \\
  \enddata
\tablecomments{Color statistics are given for two samples of galaxies in
               the HDFN: field galaxies, selected purely by \zphot\
               \citep{cap04}, and ELGs selected from
               the grism data presented here.  $N$ (columns 3 and 8)
               gives the number of galaxies with GOODS photometry in
               each sample which match the \zphot\ criteria of column
               (2) or contain the emission lines listed in column
               (7). Columns (4) and (9) give the first quartile color of
               the samples; columns (5) and (10) list the median color
               and columns (6) and (11) list the third quartile color.}
\end{deluxetable*}

\begin{figure}
\epsscale{0.85}
\plotone{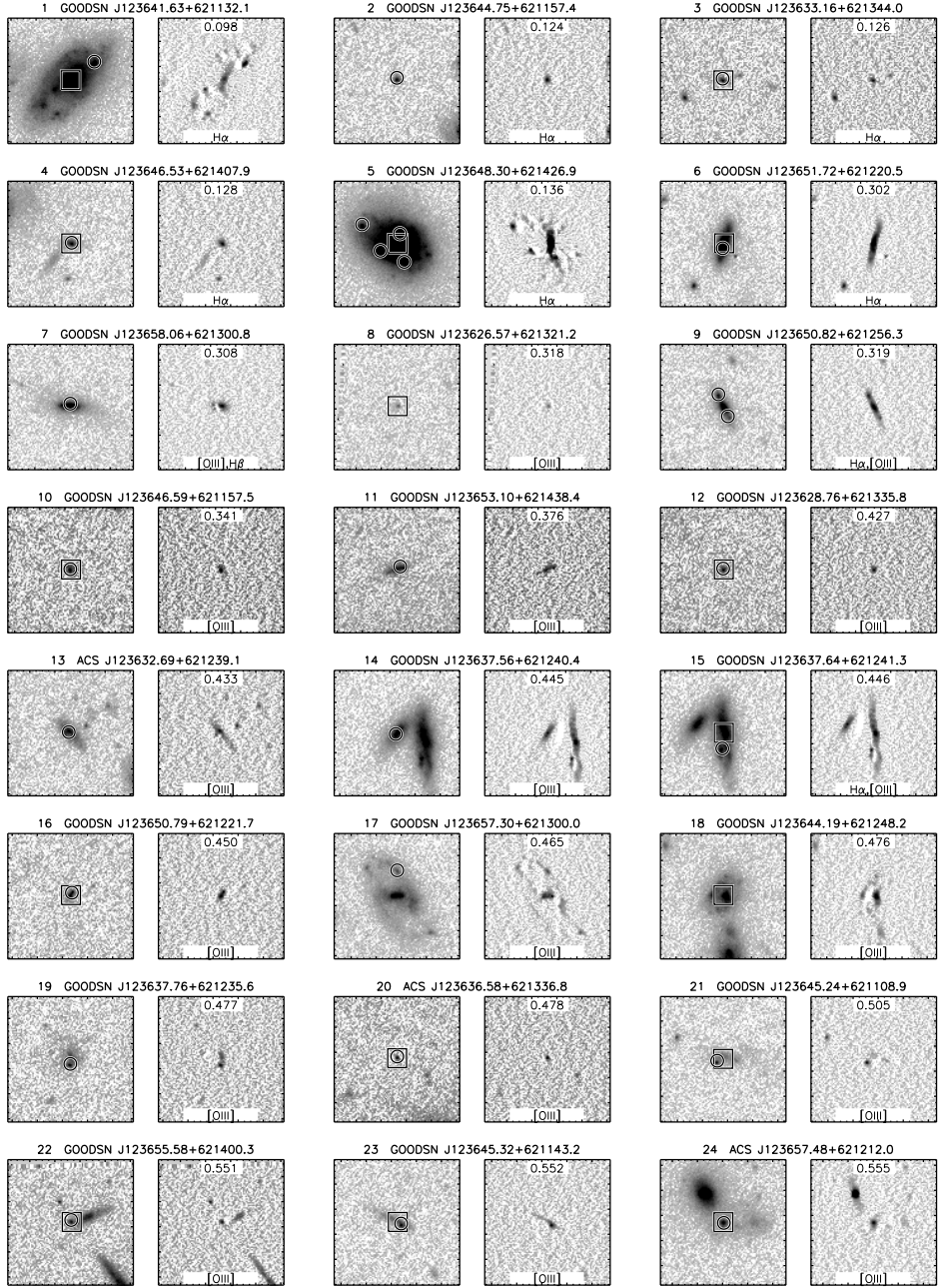}
\caption[]{\small{Postage stamp images, 5\as\ on a side, of the 46 ELGs,
  arranged by redshift.  For each galaxy, two versions of the detection
  image are shown.  The left panels show the detection image with
  spectrum extraction aperture locations containing emission lines
  marked - squares from method A, and circles from method B.  The right
  panels shows the high-pass filtered detection image (see
  Sec~\ref{ss:grsel} and Fig.~\ref{f:grexamp}) .  Here, the grism
  redshift is noted at the top of the panel, and the identified lines
  are noted at the bottom.  The final panel shows the orientation of the
  images.
  \label{f:stamps}}}
\end{figure}

\addtocounter{figure}{-1}
\begin{figure}[p]
\epsscale{0.85}
\plotone{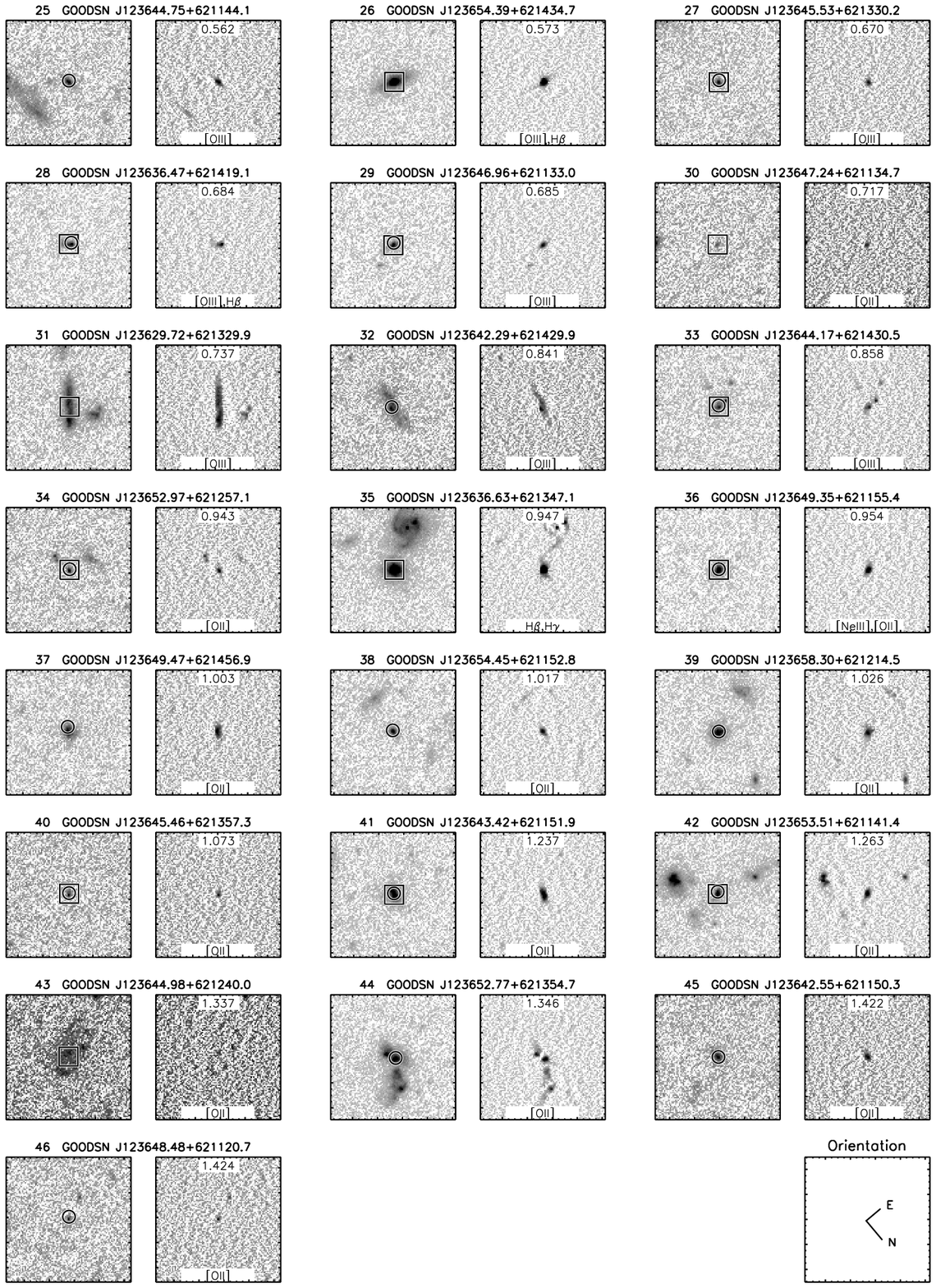}
\caption[]{Continued}
\end{figure}

\begin{figure}
\epsscale{0.7}
\plotone{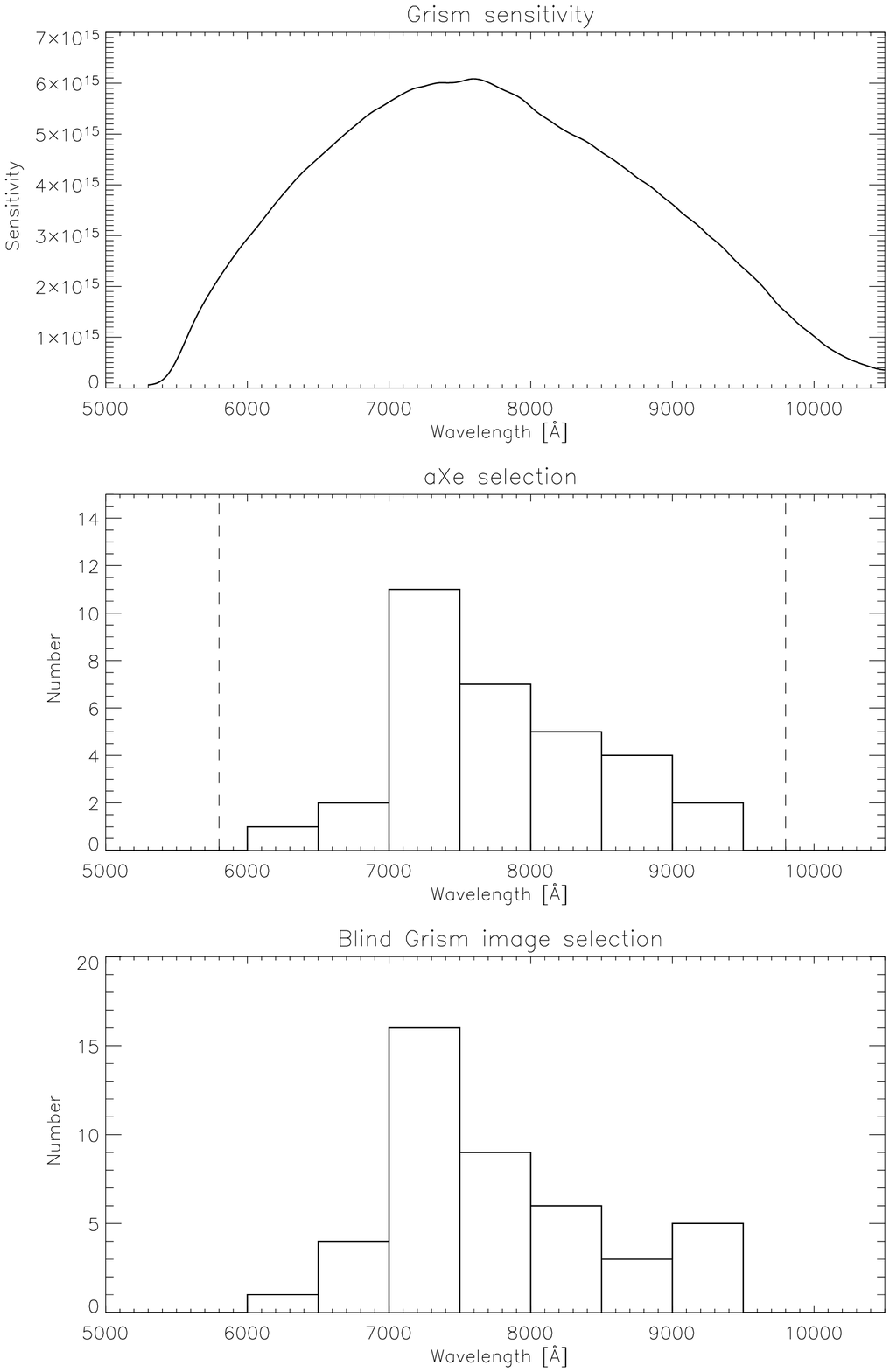}
\epsscale{1}
\caption[]{Comparison of the $\lambda$ distribution of the lines found
  in the ELGs with the grism sensitivity curve.  The grism sensitivity
  curve of \citet{wp04} is shown in the top panel.  The units of the
  ordinate are erg cm$^{-2}$ s$^{-1}$ \AA$^{-1}$ per DN s$^{-1}$. The
  middle plot shows the histogram of $\lambda$ values found with method A.
  The dashed lines delimit the $\lambda$ search range for emission
  lines.  The bottom panel shows the $\lambda$ histogram for the method
  B selected ELGs.
  \label{f:lamdist}}
\end{figure}

\begin{figure}
\epsscale{.6}
\plotone{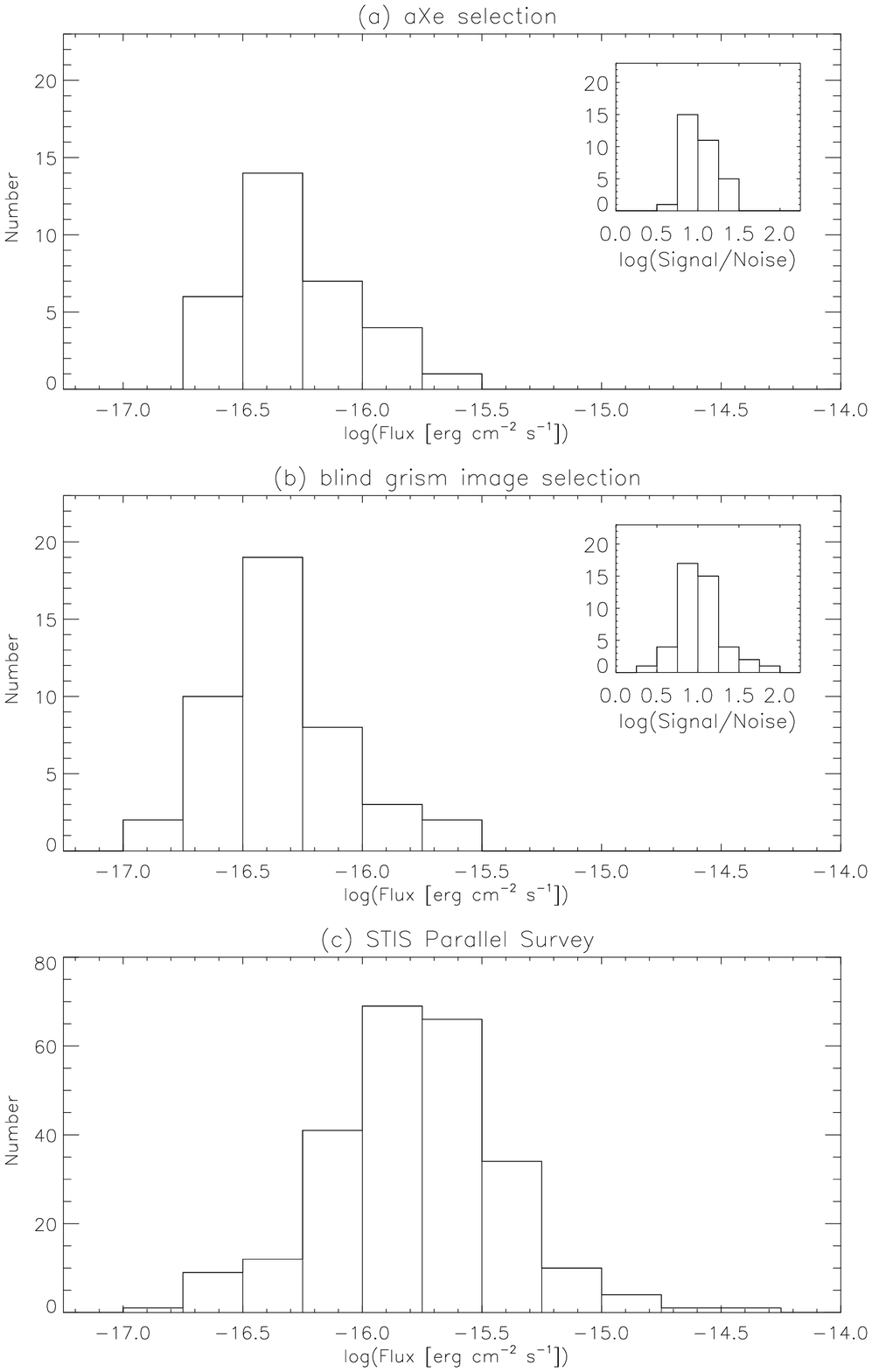}
\caption[]{Line flux distribution of the ELSs found with the direct
  image selection (panel a) and grism image selection (panel b) compared
  to the ELSs found in the STIS Parallel Survey \citep[panel
  c][]{teplitz03a}. The inset in panels a,b shows the distribution of
  the line signal-to-noise ratio.
  \label{f:flamdist}}
\end{figure}

\begin{figure}
\epsscale{.7}
\plotone{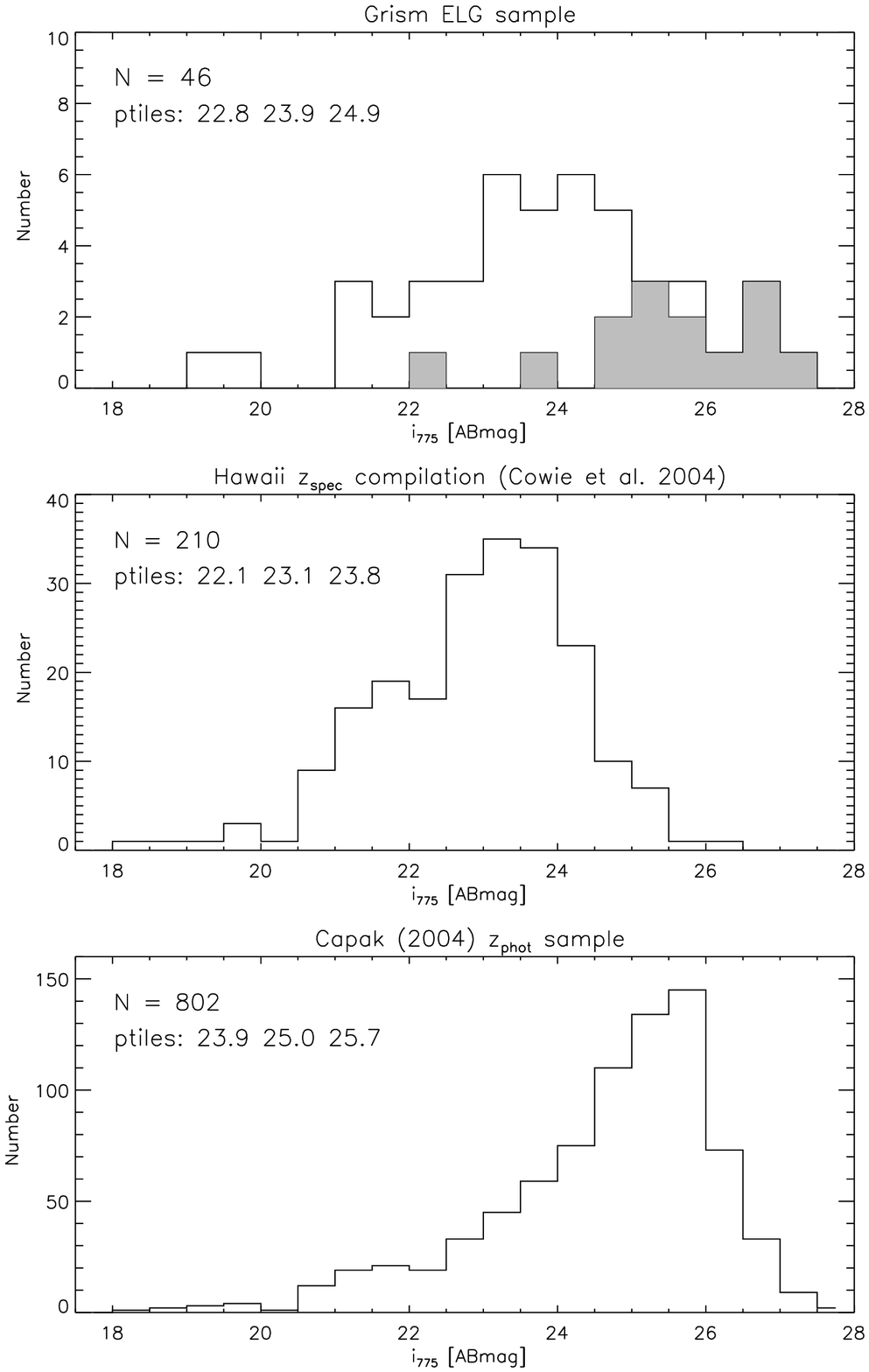}
\epsscale{1}
\caption[]{Histogram of $i_{775}$ magnitudes of the grism-selected ELG
  sample (top panel) compared with the spectroscopic redshift sample of
  \citet{cbhcs04} (middle panel) and the photometric redshift sample of
  \citet{cap04} (bottom panel).  Objects from the latter two samples are
  only included in the histograms if they match with sources cataloged
  in our detection image.  All $i_{775}$ magnitudes are derived from our
  ACS images.  In the upper left of each corner we report the total
  number of sources in the sample and the 25th, 50th (median), and 75th
  percentile $i_{775}$ magnitudes.  In the top panel sources with
  ambiguous \zgrism\ estimates (notes d-g in Table~\ref{t:source}) are
  indicated with the shaded histogram.
  \label{f:mhist}}
\end{figure}

\begin{figure}
\epsscale{0.6}
\plotone{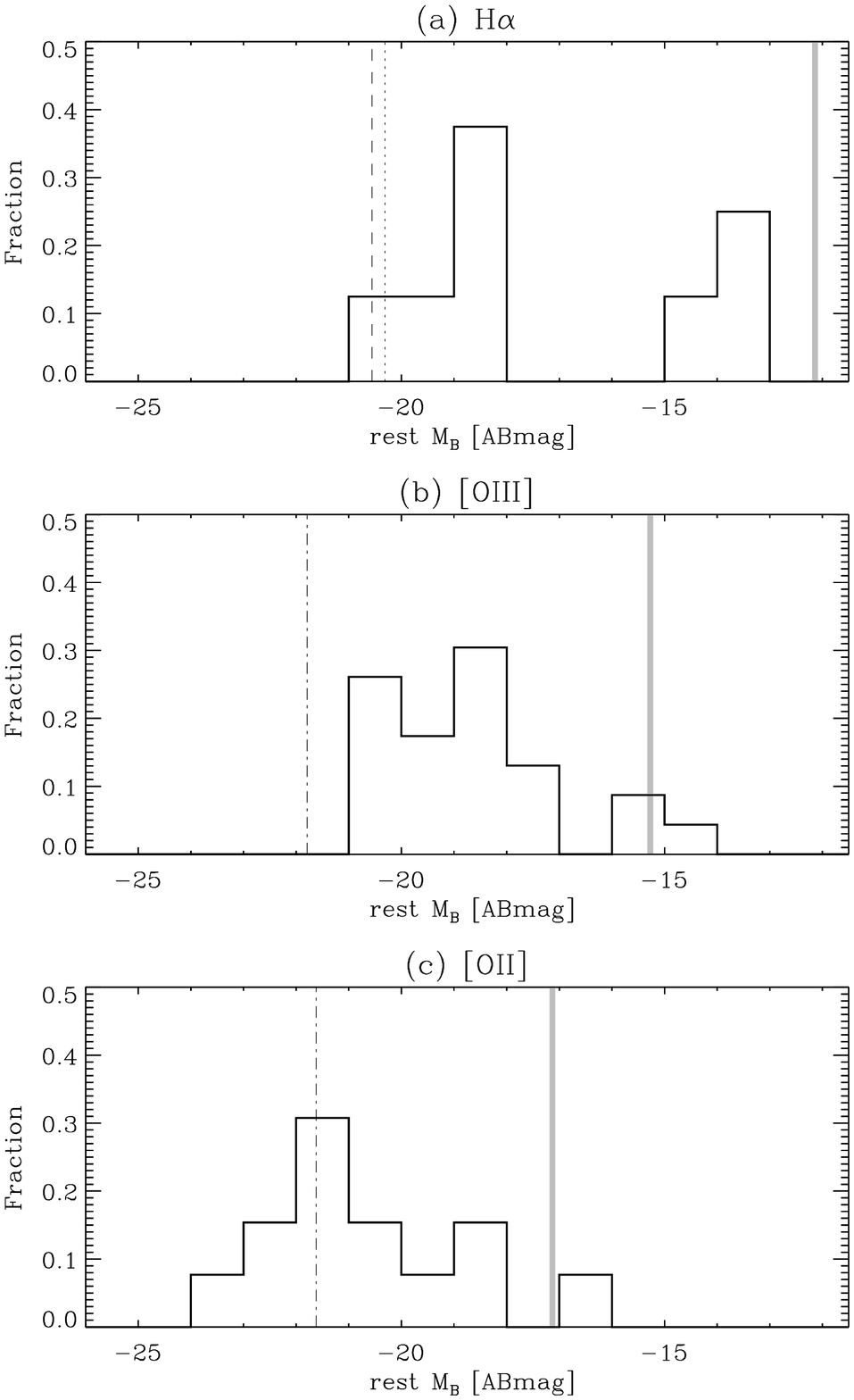}
\caption[]{Rest frame $B$ band absolute mag histograms of the the ELGs
  split by line identification, showing galaxies detected in \Halpha
  (panel a); \fion{O}{III} (panel b); and \fion{O}{II} (panel c).  The
  thick gray line shows the absolute magnitude corresponding to $m = 27$
  ABmag for a line found at $\lambda = 7500$\AA; this is a crude
  estimate of the faintest galaxies we are likely to find.  The
  broken vertical lines indicate the knee of the luminosity function,
  $M_B^\star$ of field galaxies at similar redshift to the ELG samples.
  In panel (a) the dashed line shows $M_B^\star$ derived from the Sloan
  Digital Sky Survey luminosity function \citep{blanton03} while the
  dotted line shows the $M_B^\star$ of the Two Degree Field Galaxy
  Redshift Survey \citep{madgwick02}.  The mean $z \approx 0.1$ for both
  of these surveys.  In panels (b) and (c) we show $M_B^\star$ derived
  from the $K$ band selected GOODS luminosity function at a mean $z$ of
  0.46 and 0.97, respectively \citep{dahlen05}.
  \label{f:mabshist}}
\end{figure}

\begin{figure}
\epsscale{.8}
\plotone{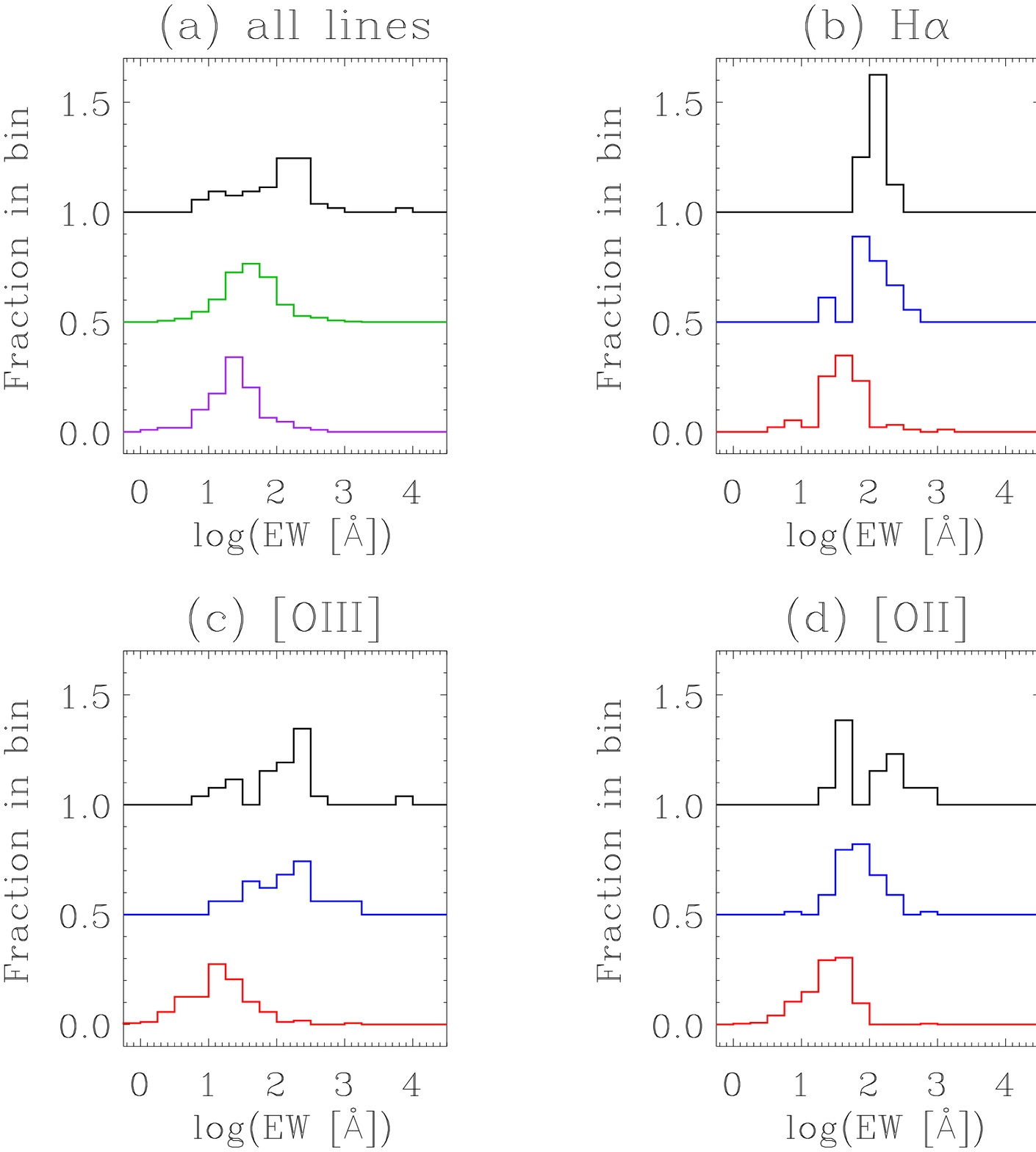}
\epsscale{1}
\caption[]{Histogram of rest frame line equivalent widths.  Histograms
  are normalized by the total sample size and offset vertically to ease
  comparison. In all cases our results are shown as the thick black line
  at top  In panel (a) we compare the widths of all lines with two local
  samples: the KISS red surveys \citep{kissr1,kissr2}, shown in green (middle),
  and the SINGG survey \citep{singg1}, shown in red (bottom).  The
  remaining panels (b,c,d) split the sources by line identification and 
  our sample is compared to two surveys that
  extend out to moderate redshifts: the STIS Parallel Survey
  \citep{teplitz03a} shown with the blue middle line and the
  Canada-France Redshift Survey \citep{cfrs14} shown with the red line
  at bottom.  
  \label{f:ewhist}}
\end{figure}

\begin{figure}
\plotone{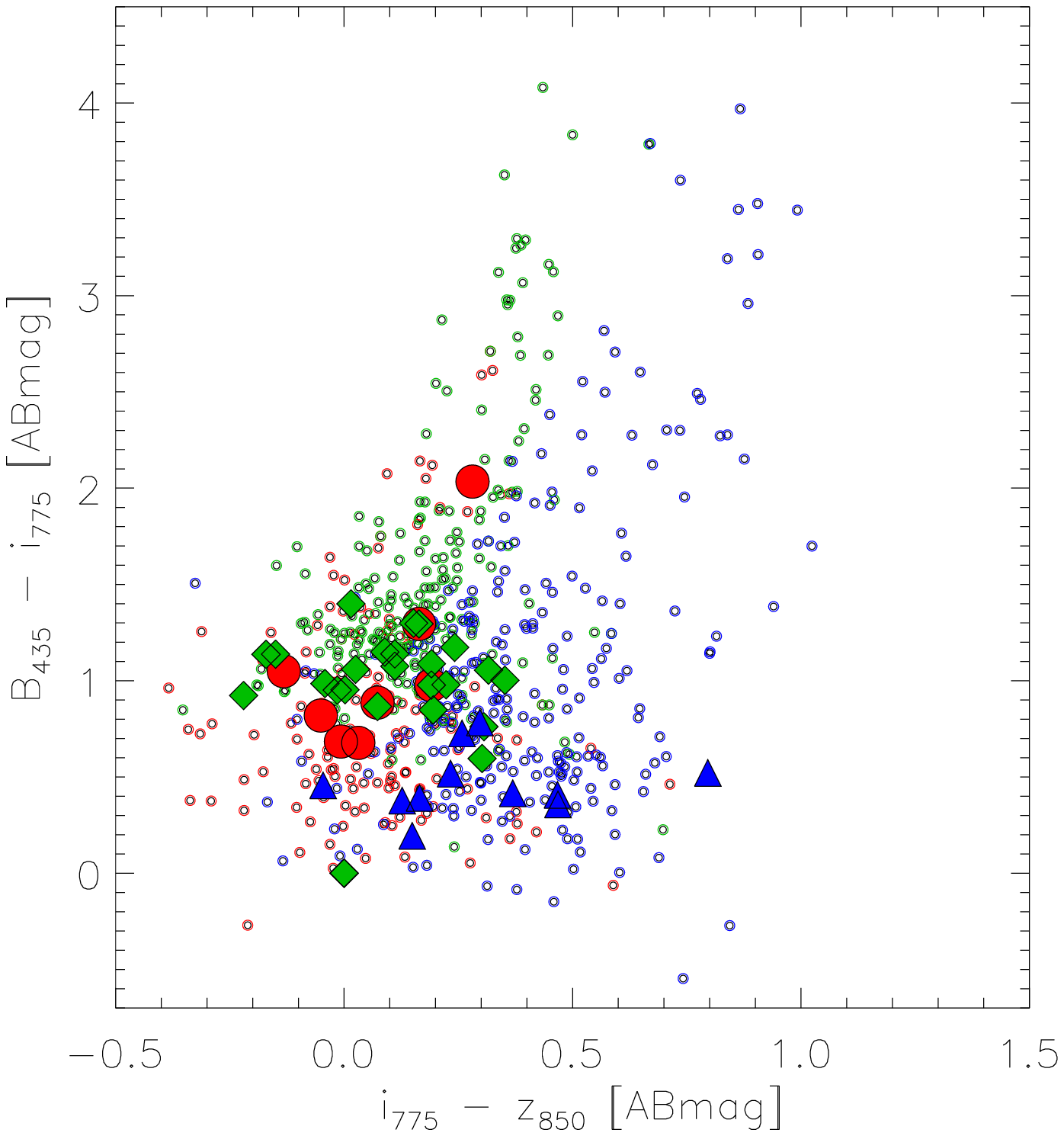}
\caption[]{The $B_{435} - i_{775}$ versus $i_{775} - z_{850}$ two color
  plot of sources in the HDFN, using GOODS photometry \citep{goods04a}.
  Small circles are sources with $\zphot \leq 1.62$ \citep{cap04},
  filled symbols are ELGs identified in this study.  The symbols are
  color coded by \zphot\ and line identification respectively, where red
  corresponds to $\zphot < 0.30$ and \Halpha\ emitters (circles), green
  corresponds to $0.3 \leq \zphot < 0.76$ and \fion{O}{III} emitters
  (diamonds), and blue corresponds to $0.76 \leq \zphot\ < 1.62$ and
  \fion{O}{II} emitters (triangles).  
  \label{f:biiz}}
\end{figure}

\begin{figure}
\plotone{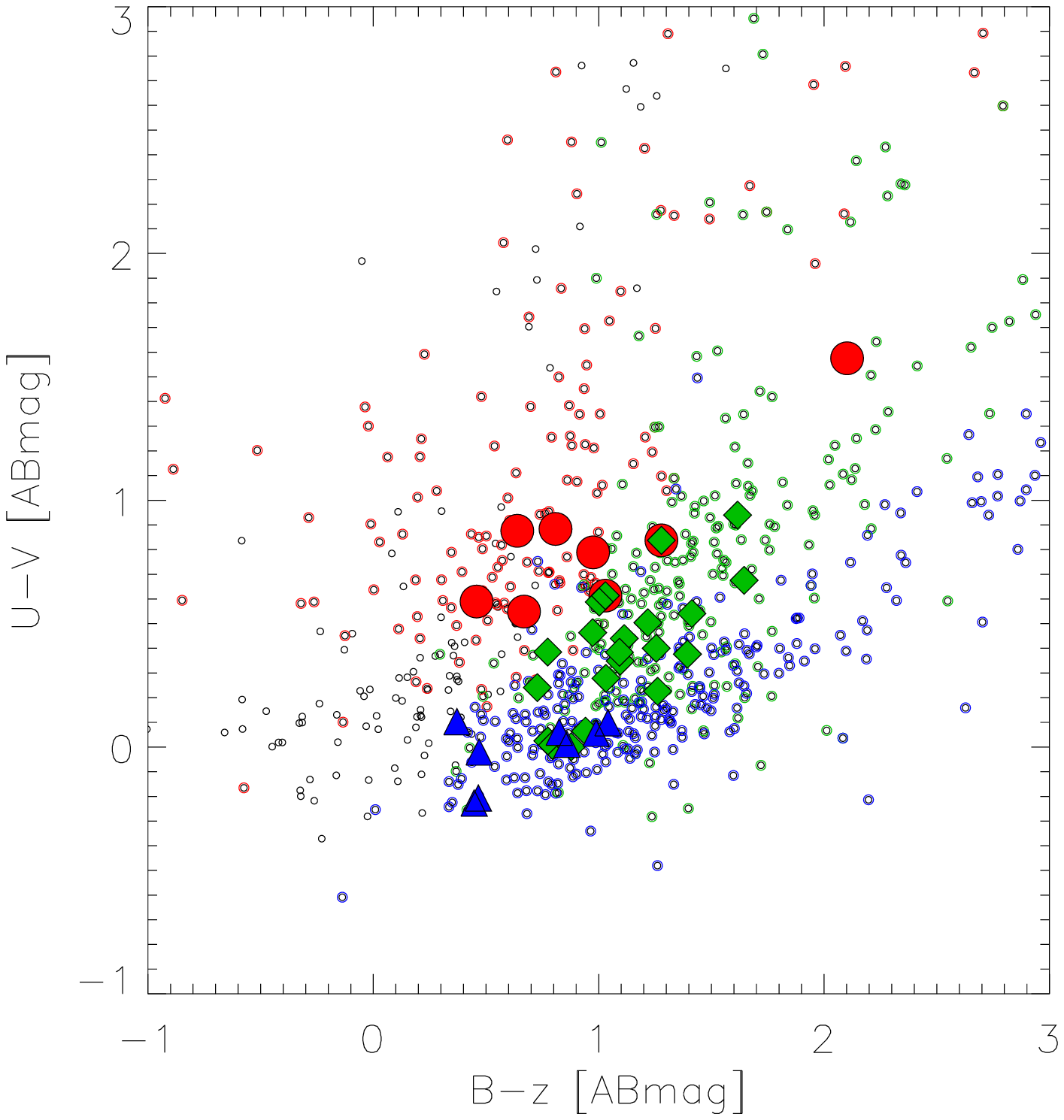}
\caption[]{The $(U-V)$ versus $(B-z)$ photometry for sources in the HDFN
  using the Hawaii group photometry \citep{cap04}.  The line
  identifications are fairly well sorted in this plane.  The symbols
  used are the same as in Fig.~\ref{f:biiz}.
  \label{f:uvbz}}
\end{figure}

\begin{figure}
\plotone{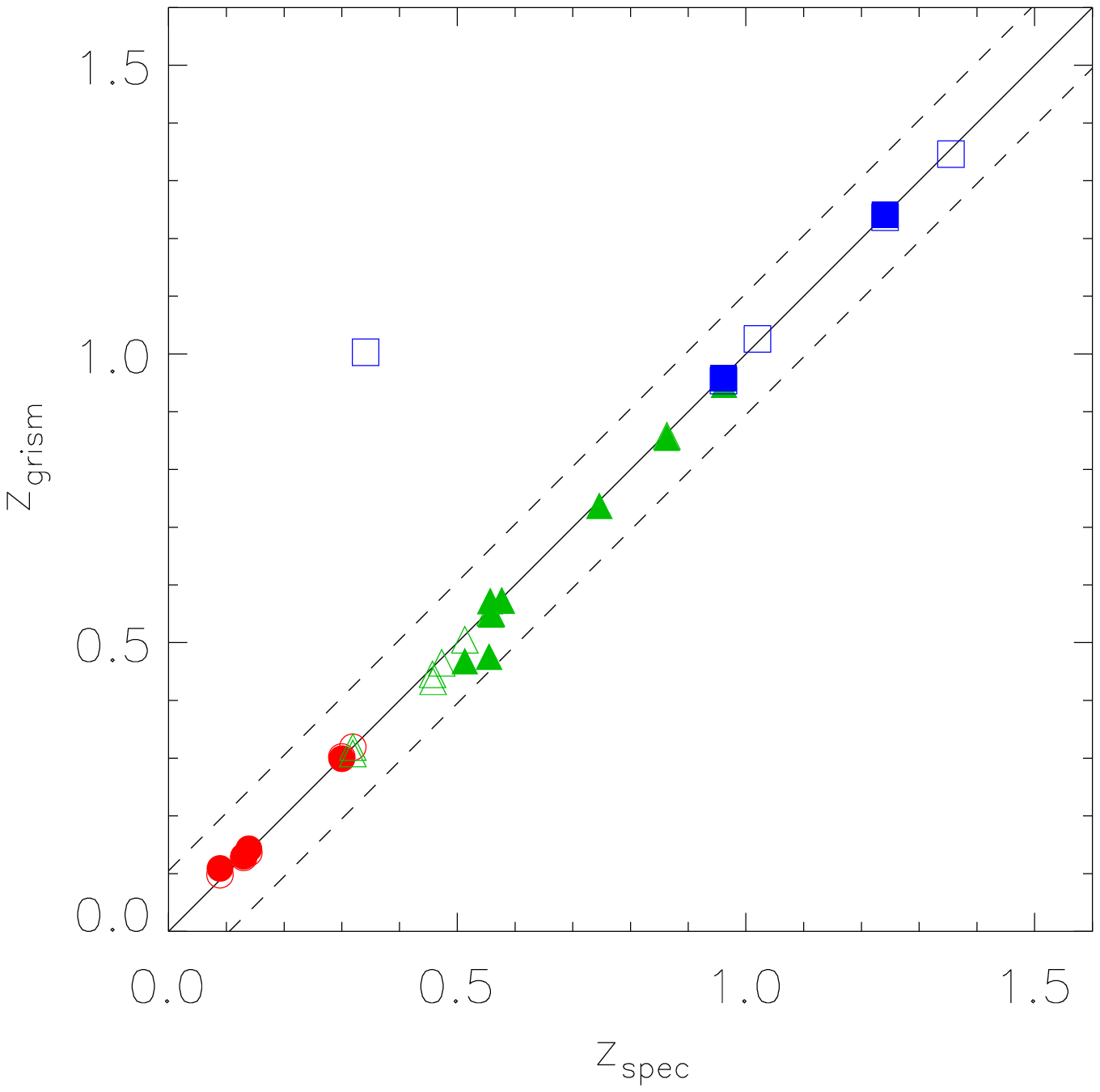}
\caption[]{Comparison of spectroscopic redshifts from \citep{cbhcs04}
  and grism redshifts from this work.  Measurements from method A are
  shown with solid symbols, measurements from method B are shown as open
  symbols.  The symbol shape and color indicate the grism line
  identification: \Halpha\ emitters are red circles, \fion{O}{III}
  emitters are green triangles, and \fion{O}{II} emitters are blue
  squares.  The unity relationship is shown as a solid line, sources
  outside the dashed lines at $\Delta z = \pm 0.105$ are considered
  outliers.
  \label{f:zgrzsp}}
\end{figure}

\begin{figure}
\plotone{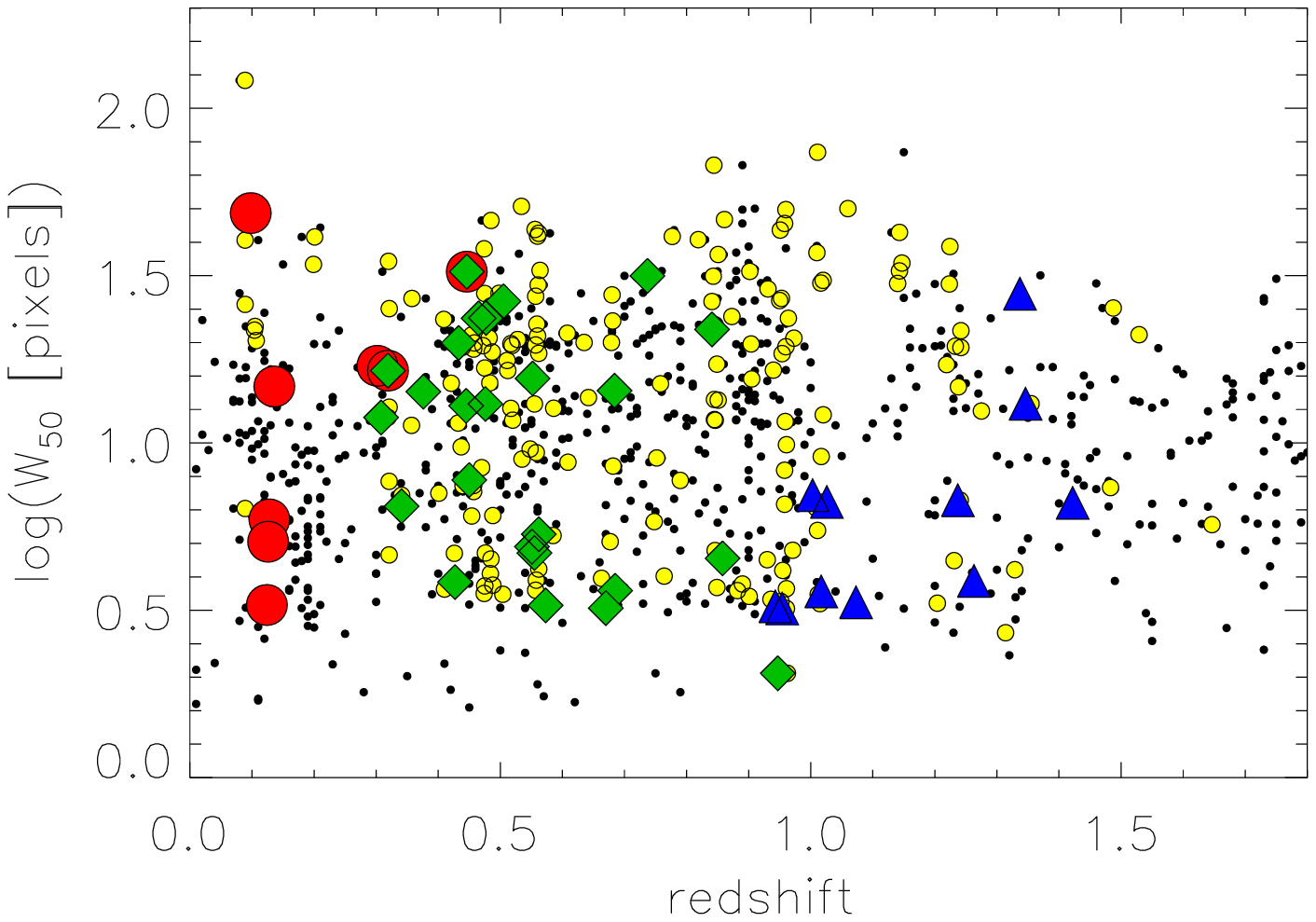}
\caption[]{Angular size plotted against redshift for the sources in our
  field.  Sources with photometric redshifts are plotted as small black
  dots.  Yellow filled circles mark sources with spectroscopic
  redshifts.  The large color filled symbols mark the sources with grism
  redshifts; red filled circles indicate \Halpha\ emitters, green
  diamonds indicate \fion{O}{III} or \Hbeta\ emitters, and blue
  triangles indicate \fion{O}{II} emitters.
  \label{f:sizz}}
\end{figure}

\end{document}